\documentclass[fleqn,usenatbib]{mnras}

\usepackage{newtxtext,newtxmath}

\usepackage[T1]{fontenc}
\usepackage{ae,aecompl}

\renewcommand{\exp}[1]{\ensuremath{{\rm exp}\left(#1\right)}}

\usepackage{graphicx}	
\usepackage{amsmath}	
\usepackage{amssymb}	
\usepackage[figure,figure*]{hypcap}

\title[EoR Power Spectra from Four Seasons of the MWA]{Deep multi-redshift limits on Epoch of Reionisation 21~cm Power Spectra from Four Seasons of Murchison Widefield Array Observations}

\author[Trott et al.]{Cathryn M. Trott$^{1,2, \dagger}$\thanks{cathryn.trott@curtin.edu.au},
C.~H. Jordan$^{1,2}$,
S.~Midgley$^{3}$,
N.~Barry$^{8,2}$,
B.~Greig$^{8,2}$,
\newauthor
B.~Pindor$^{8,2}$,
J.~H.~Cook$^{1}$,
G.~Sleap$^{12}$,
S.~J.~Tingay${^1}$,
D.~Ung$^{1}$,
P.~Hancock${^1}$,
\newauthor
A.~Williams$^{12}$,
J.~Bowman$^{13}$,
R.~Byrne$^{6}$,
A.~Chokshi$^{8,2}$,
B.~J.~Hazelton$^{6,9}$,
K.~Hasegawa$^{5}$,
\newauthor
D.~Jacobs$^{13}$,
R.~C.~Joseph$^{1,2}$,
W.~Li$^{11}$,
JLB~Line$^{1,2}$,
C.~Lynch$^{1,2}$,
B.~McKinley$^{1,2}$,
\newauthor
D.~A.~Mitchell$^{7}$,
M.~F.~Morales$^{6,2}$,
M.~Ouchi$^{15,16,17}$,
J.~C.~Pober$^{10}$,
M.~Rahimi$^{8,2}$,
\newauthor
K.~Takahashi$^{4,14}$,
R.~B.~Wayth$^{1,2}$,
R.~L.~Webster$^{8,2}$,
M.~Wilensky$^{6}$,
J.~S.~B.~Wyithe$^{8,2}$,
\newauthor
S. Yoshiura$^{8,2}$,
Z.~Zhang$^{10}$,
Q.~Zheng$^{11}$\\
$^1$International Centre for Radio Astronomy Research (ICRAR), Curtin University, Bentley WA, Australia\\
$^2$ARC Centre of Excellence for All Sky Astrophysics in 3 Dimensions (ASTRO 3D), Bentley, Australia\\
$^{3}$DownUnder GeoSolutions Pty Ltd, West Perth, Australia\\
$^4$Faculty of Advanced Science and Technology, Kumamoto University, Japan\\
$^{5}$Graduate School of Science, Nagoya University, Japan\\
$^6$Department of Physics, University of Washington, Seattle, WA 98195, USA\\
$^{7}$CSIRO Astronomy \& Space Science, Australia Telescope National Facility, P.O. Box 76, Epping, NSW 1710, Australia\\
$^{8}$School of Physics, The University of Melbourne, Parkville, VIC 3010, Australia\\
$^{9}$University of Washington, eScience Institute, Seattle, WA 98195, USA\\
$^{10}$Brown University, Department of Physics, Providence, RI 02912, USA\\
$^{11}$Shanghai Astronomical Observatory, China\\
$^{12}$Curtin Institute of Radio Astronomy, GPO Box U1987, Perth, WA 6845, Australia\\
$^{13}$Arizona State University, Tempe, AZ, USA\\
$^{14}$International Research Organization for Advanced Science and Technology, Kumamoto University, Japan\\
$^{15}$Division of Science, National Astronomical Observatory of Japan, 2-21-1 Osawa, Mitaka, Tokyo 181-8588, Japan\\
$^{16}$Institute for Cosmic Ray Research, The University of Tokyo, 5-1-5 Kashiwanoha, Kashiwa, Chiba 277-8582, Japan\\
$^{17}$Kavli Institute for the Physics and Mathematics of the Universe (Kavli IPMU, WPI),\\ The University of Tokyo, 5-1-5 Kashiwanoha, Kashiwa, Chiba, 277-8583, Japan\\
$^\dagger$ARC Future Fellow}
\date{Accepted 2020 February 5. Received 2020 January 28; in original form 2019 December 15.}

\pubyear{2020}

\begin{document}
\label{firstpage}
\pagerange{\pageref{firstpage}--\pageref{lastpage}}
\maketitle


\begin{abstract}
We compute the spherically-averaged power spectrum from four seasons of data obtained for the Epoch of Reionisation (EoR) project observed with the Murchison Widefield Array (MWA). We measure the EoR power spectrum over $k= 0.07-3.0~h$Mpc$^{-1}$ at redshifts $z=6.5-8.7$. The largest aggregation of 110 hours on EoR0 high-band (3,340 observations), yields a lowest measurement of (43~mK)$^2$ = 1.8$\times$10$^3$ mK$^2$ at $k$=0.14~$h$Mpc$^{-1}$ and $z=6.5$ (2$\sigma$ thermal noise plus sample variance). Using the Real-Time System to calibrate and the CHIPS pipeline to estimate power spectra, we select the best observations from the central five pointings within the 2013--2016 observing seasons, observing three independent fields and in two frequency bands. This yields 13,591 2-minute snapshots (453 hours), based on a quality assurance metric that measures ionospheric activity. We perform another cut to remove poorly-calibrated data, based on power in the foreground-dominated and EoR-dominated regions of the two-dimensional power spectrum, reducing the set to 12,569 observations (419 hours). These data are processed in groups of 20 observations, to retain the capacity to identify poor data, and used to analyse the evolution and structure of the data over field, frequency, and data quality. We subsequently choose the cleanest 8,935 observations (298 hours of data) to form integrated power spectra over the different fields, pointings and redshift ranges.
\end{abstract}




\section{Introduction}
The Epoch of Reionisation (EoR) marks a period of remarkable change in the Universe, witnessing the heating and ionising of the neutral hydrogen that filled the intergalactic medium, via the ultraviolet photons from the first generations of stars and their remnants \citep{furlanetto06}. While the integrated information provided by the Thompson scattering effects on Cosmic Microwave Background (CMB) photons, and line-of-sight information provided by the IGM path to high-redshift quasars, galaxies and gamma-ray bursts \citep{fan06,yang19,jiang16,ouchi10,totani06} offer clues and constraints on the spatial and redshift evolution of this period, direct study of the neutral hydrogen signal via its radio hyperfine transition at $\lambda_{\rm rest}=21$~cm provides one of the best observational tracers because it can provide redshift-dependent and spatially-dependent information, and is isotropic and ubiquitous \citep{bowman09}. Recently, Bowman and colleagues reported the detection of an absorption trough in low radio frequency globally-averaged sky power, which they identified with the Cosmic Dawn, preceding the EoR, wherein the light from the first generations of stars coupled the hydrogen spin temperature to the gas kinetic temperature, providing contrast to the CMB photon temperature \citep{bowman18}. This detection provided a globally-averaged (all sky) signpost for the further evolution of the Universe, but does not provide the spatial information required to estimate the underlying astrophysical parameters of interest that characterise the properties of the first stars and galaxies, and the IGM gas. For that, interferometric measurements with low-frequency radio telescopes can provide the spatial information.

The initial detection, and future exploration of the EoR, are therefore primary experiments for low-frequency radio telescopes sensitive to the redshifted emission, such as the MWA \citep{tingay13_mwasystem,bowman13_mwascience,wayth18}, the Low-Frequency Array (LOFAR){\footnote[1]{http://www.lofar.org}} \citep{vanhaarlem13}, the Precision Array for Probing the Epoch of Reionization (PAPER){\footnote[2]{http://eor.berkeley.edu}} \citep{parsons10}, the Long Wavelength Array (LWA){\footnote[3]{www.lwa.unm.edu}} \citep{taylor07}, and the upcoming Hydrogen Epoch of Reionization Array (HERA){\footnote[4]{http://reionization.org}} \citep{deboer17} and Square Kilometre Array (SKA){\footnote[5]{http://skatelescope.org}} \citep{koopmans15}.

Progress in the field has been hampered by the systematic contamination of the signal caused by inaccurate and imprecise calibration, and spectrally-structured foreground signals from radio galaxies and Galactic emission. Over the past five years there has been a wealth of research undertaken to improve the data treatment methods to mitigate the systematics, including the calibration model \citep{barry16,trottwayth16,offringa15,patil17,procopio17,ewallwice17,orosz19,kern19,dillon18}, instrument model \citep{li18,joseph18,eloy17,trott17}, power spectrum methodology \citep{offringa19,barry19,trott16,choudhuri17,parsons10,parsons12,liu14}, foregrounds \citep{datta10,vedantham12,trott12,thyagarajan15a,thyagarajan15b,chapman14,eastwood18,mertens18}, and ionospheric effects \citep{jordan17,trott18,mevius16}. This concerted effort and broad approach have paved the way for the current datasets to be used for EoR science.

Currently, LOFAR, LWA and MWA are contributing ongoing and dedicated effort to analyse the thousands of hours of data collected by their experiments, and are publishing results from the best of these data. Recent reported measurements include those of \citet{patil17} and \citet{gehlot18} from LOFAR, \citet{eastwood19} from LWA, and \citet{barry19a} and \citet{li19} from MWA. LOFAR have deep observations in two fields (NCP and 3C196), although all published work uses the NCP field only. The PAPER array has also replaced all of their previous results with a robust re-analysis of their data \citep{cheng18,kolopanis19}. However these reports have often used relatively small sets of data, obtained over a given observing field and in a limited time duration. In this work, we use data quality metrics to assess the quality of 13,591 MWA observations (453 hours) observed from August 2013 to January 2017 over three observing fields and in two bands. These bands span 139 -- 197~MHz, corresponding to $z=9.3-6.2$, a time when the EoR signal is expected to be observed in emission with respect to the CMB, and decreasing in power with decreasing redshift. We present multi-redshift limits from the largest set of data ever aggregated, moving from the sets of tens of hours towards the thousand-hour nominal dataset required to yield a detection. The breadth and depth of the datasets provides a stringent set of results that set the path forward for the MWA experiment and future SKA.

In Section \ref{Sec:methods}, we review the methods used to form power spectra, before describing the observations, datasets, data quality metrics, and simulations to ensure no signal loss. Section \ref{sec:results} then presents the results for each field and redshift, before the best sets are combined to form the final upper limits on the signal power. We then compare the different fields in Section \ref{sec:comparison} before discussing next steps.

\section{Methods}\label{Sec:methods}
\subsection{Power spectrum methodology}
The spatial power spectrum quantifies the signal power as a function of spatial scale, $k$ ($h$Mpc$^{-1}$). It is the Fourier Transform of the two-point spatial correlation function, and can be estimated from the volume-normalised Fourier-Transformed brightness temperature field:
\begin{equation}
    P(|\vec{k}|) = \int_V \xi(\vec{r})\exp{(-2\pi{i} \vec{k}\cdot\vec{r})} d\vec{r} = \frac{1}{\Omega} \langle \tilde{T}(k)^\dagger \tilde{T}(k) \rangle.
\end{equation}

In radio interferometry, the angular scales are related to the Fourier modes of the measured interferometric visibilities, and the line-of-sight modes can be mapped with spectral channels (for a resonant line signal): $(u,v) \rightarrow k_\bot$, $\mathcal{F}(f) = \eta \rightarrow k_\parallel$. As such, we extract angular modes directly from the measured visibilities, such that:
\begin{eqnarray}
    P(|\vec{k}|) &=& \frac{1}{\Omega} \langle \tilde{V}(k)^\dagger \tilde{V}(k) \rangle\\\nonumber
    V(\vec{k}=(u,v,\eta)) &=& \iint_{Af} \frac{2k_BT}{\lambda^2} \exp{-2\pi{i}(ul+vm+f\eta)} dA df,
\end{eqnarray}
where $A$ is the angular dimension and $\Omega$ is the observing volume. The brightness temperature and source flux density are related via the equation for the specific intensity, which is linear in the radio regime:
\begin{equation}
    S = 10^{26}\frac{2k_BT}{\lambda^2} \,\, {\rm Jy\,\,sr}^{-1}
\end{equation}
where $k_B$ is the Boltzmann constant. During the Fourier Transform, the area dimension is collected to yield power spectral flux density (Jansky).

Algorithmically, the power spectrum is formed from a data cube with dimensions $(u,v,f)$, where the angular Fourier Transform to ($u,v$) is performed natively by the interferometer, and the spectral channels are used to map the line-of-sight. Data are observed over time and integrated together coherently by gridding measured visibilities onto a common discretized $uv$-plane.
The final steps are to Fourier Transform along frequency in each cell, and then to square to arrive at the unnormalised power. A power spectrum formed in this way may be used for cosmological measurements, because the three $k$-vectors are orthogonal and can easily be mapped to spherical $k$. In this work, the 2D (cylindrically-averaged) and 1D (spherically-averaged) power spectrum are used. The former principally acts to identify foreground leakage into the parameter space used for EoR analysis. The latter provides the cosmologically-relevant measurements. In 2D, the angular and line-of-sight modes are separated, and denoted $k_\bot, k_\parallel$. In 1D, these are averaged to $k^2 = k_\bot^2 + k_\parallel^2$. We present the dimensionless power spectrum in 1D:
\begin{equation}
    \Delta^2(k) = \frac{k^3P(k)}{2\pi^2} \,\,\rm{mK}^2.
\end{equation}

The alternative approach to approximating the power spectrum is the delay transform \citep{parsons12,jacobs14}. Here, each individual baseline's data are Fourier Transformed along its frequency channels, and the power formed through their squared quantities. This approach is straight forward, but can be difficult to interpret cosmologically, because the $k$-vectors are non-orthogonal except on short baselines (the line-of-sight transform evolves with frequency) and there is no angular correlation encoded between baselines with similar lengths and orientations. Nonetheless, the delay transform is very useful for quality assurance to ensure the visibility data are not corrupted or contaminated in the regions of parameter space used for EoR analysis (i.e., the `EoR Window', a region of $k_\bot,k_\parallel$ parameter space outside of the region expected to be dominated by smooth-spectrum foregrounds).

CHIPS -- the Cosmological HI Power Spectrum estimator \citep{trott16} -- is one of the signal processing pipelines used by the MWA EoR collaboration for taking calibrated data and processing them to output power spectra, with associated uncertainties. In its original form, it was intended to undertake a full thermal noise plus residual foreground signal inverse covariance weighting, to optimally extract cosmological information. There are difficulties with this approach, and these were explored in the literature at the time, and have more recently been demonstrated in the retracted results from the PAPER collaboration \citep{cheng18}. Inaccurate residual foreground models, and failure to fully and independently understand their internal covariance and covariance with the signal, can easily lead to signal loss. As such, CHIPS is used primarily (and entirely in this work), as an inverse variance estimator, where the baseline weighting is used for sampling.

Line-of-sight modes for computing the power spectrum are extracted from spectral sampling of the data. The MWA's signal processing chain contains filterbanks that yield 24 coarse channels of 1.28~MHz over the full 30.72~MHz band. Within these coarse channels, the native spectral resolution is 10~kHz, but EoR data are observed at 40~kHz resolution and processed at 80~kHz resolution. The shape of the fine polyphase filterbank leads to poor bandpass characteristics at the coarse channel edges, and as such, a single 80~kHz channel is flagged at each end of each coarse channel. This yields regularly-spaced missing channels in the final output visibilities. A Fourier Transform over the data to retrieve the line-of-sight spatial scales will contain a comb shape due to these missing channels, where the $k_\parallel=0$ mode is copied in harmonics of the coarse channel separation. There are several ways to handle this; in this implementation of CHIPS, we use an ordinary kriging (a Gaussian Process Regression) to provide an interpolate of these data that uses the covariance structure of the data \citep{rasmussen96,wackernagel}. Kriging has been used to fit for foregrounds in LOFAR EoR datasets, using an optimised set of hyperparameters \citep{mertens18}. The kriging kernel (variogram) is estimated conservatively to contain a noise-like variance and a frequency covariance which decays smoothly across several megahertz. Kriging applies an interpolation over unsampled data using linear weights of sampled data. The weights are computed to minimise the mean-squared error and yield an unbiased interpolate, subject to a specified data covariance matrix. In this work, we assume a covariance matrix with a nugget (variance term) provided by the thermal noise variance, and a Gaussian-shaped spectral covariance for the foregrounds with a characteristic length of 50 channels. The functional form of the kriging kernel is kept consistent across all datasets to avoid biasing results with fine-tuning, and is given by:
\begin{equation}
    K(\nu,\nu^\prime) = 0.1\delta(\nu-\nu^\prime) + \exp{-\frac{(\nu-\nu^\prime)^2}{4\rm{MHz}^2}},
\end{equation}
where the relative scaling of the nugget to the squared-exponential is appropriate for the relative amplitude of the thermal noise and foregrounds. The same kriging kernel is applied to all datasets, and is applied to the real and imaginary components of each $uv$-cell along the frequency distribution in the gridded $(u,v,\nu)$ data cube.
The results are not strongly-dependent on the choice of spectral scale, with statistically-similar results occurring for values of 2--5 MHz. Testing demonstrates that the application of kriging does not bias results in the modes used for measurement (see Section \ref{sec:simulations}), but can at higher-order modes. It was initially implemented to access more modes close to the coarse channel harmonics, but this has provided only limited improvement for some datasets. Nonetheless, it does yield improved results in those modes used for limits in this work ($k<0.4$ $h$Mpc$^{-1}$) compared with omitting the kriging. In future work, we will either (1) not apply kriging at all; (2) invest more effort to understand and refine it so that we are confident that is unbiased across the full range of $k$-modes. In this work, we retain it, because it is infeasible to re-process all of these data and we report results at unbiased wave modes only.

\subsection{Observations}
Data were observed with the Murchison Widefield Array, a general-purpose low-frequency radio interferometer operating at the Murchison Radioastronomy Observatory in Western Australia \citep{tingay13_mwasystem,bowman13_mwascience}. Phase I of the array comprised a pseudo-random core for EoR science, surrounded by sparser remote tiles for angular resolution. In Phase II of the array \citep{wayth18,beardsley19}, the telescope consists of 256 tiles of 16 dual-polarisation dipole antennas in a regular 4$\times$4 grid, spread over $\sim$5~km. Only 128 tiles can be connected to the signal chain at one time, and the telescope operates in an "Extended" (survey science; long baselines) or "Compact" (high surface brightness sensitivity; short baselines) configuration. The EoR experiment uses the latter configuration. The sky model for instrument calibration and source subtraction are formed from the Phase I configuration, and auxiliary data from other telescopes.

The EoR experiment observes in three bands \citep{jacobs16}: ultralow-band (75--100~MHz), low-band (139--167~MHz), and high-band (167--197~MHz), with more than ninety percent of data observed in low- and high-bands. The data for this work are only taken from the upper two bands (139--197~MHz), consistent with the reionisation epoch when the signal is expected to be in emission (as compared with the Cosmic Dawn). The primary EoR experiment observes data from three observing fields, chosen to minimise sky temperature (away from the Galactic plane) and containing bright calibration sources. They are EoR0 (RA=0h, Dec=$-$27$^o$), EoR1 (RA=4h, Dec=$-$27$^o$), and EoR2 (RA=10.3h, Dec=$-$10$^o$). EoR1 contains Fornax A, an extended ($\sim$1 degree) double-lobed radio source part way down the main primary beam, and EoR2 contains Hydra A. These two sources both help and hinder data calibration and need to be subtracted with high precision for EoR science \citep{procopio17,trottwayth16}. EoR0 contains the setting Galactic plane in early (pre-zenith) pointings, yielding power from the horizon in power spectra. Of the data presented in this paper, 50\% are EoR0, 28\% are EoR1, and 22\% are EoR2. This is a function of the data that have been calibrated, and not a reflection on the overall contributions of each. However, in general EoR2 is observable at the end of the season, and there are fewer hours available for it than for EoR0 and EoR1.

The data used in this work span Phase I and II of the array. Despite redundancy being available in Phase II, and used in other MWA pipelines such as Fast Holographic Deconvolution \citep[FHD,]{barry19} plus Omnical \citep{li18,li19,zheng20}, the RTS calibration currently only performs sky-based calibration. Given the interest in understanding the utility of hybrid arrays with redundant and non-redundant baselines, it would be useful to compare the results for the different configurations. Unfortunately, the data in this work are 92\% Phase I and 8\% Phase II, with no individual Phase II set exceeding 5 hours. As such, comparison of datasets suffers from the small and concentrated number of observations, and is not useful.

Data used in this work were observed over five pointings, separated by $\sim$6.8 degrees on the sky (27 minutes per pointing), corresponding to the beamformer analogue delay settings that produce a consistent primary beam response. Of all of the pointings, the five central pointings (including zenith) are found to have the least contaminated power \citep{beardsley16}, and have the most well-behaved beam patterns, and are used exclusively in this work. Off-zenith pointings are only pursued for EoR0 and EoR2, because the zenith results from these fields are sufficiently-interesting to warrant coherently combining data from differing pointings. Table \ref{table:pointings} lists the Alt/Az and name for each pointing used here.
\begin{table}
\centering
\begin{tabular}{|c||c|c}
\hline 
Name & Altitude (deg) & Azimuth (deg) \\
\hline
Minus2 & 76.3 &  90\\
Minus1 & 83.2 & 90 \\
Zenith & 90 & 0 \\
Plus1 & 83.2 & 270 \\
Plus2 & 76.3 & 270 \\
\hline
\end{tabular}
\caption{Names and sky locations of the five pointings of the beam used in this work.}\label{table:pointings}
\end{table}

The data for this work were observed in the 2013B, 2014B, 2015B, 2016B observing seasons, with a range of August 2013 -- January 2017, spanning Phase I and Phase II configurations.

The data were selected by extracting all of the observations available in the MWA database (hosted by the Pawsey Supercomputing Centre) from 2013 to 2017, which had been successfully calibrated with the RTS (output files were produced with finite values, and bandpass and phase plots looked reasonable), had complete calibrated visibility files with the standard temporal (8 second) and spectral (80~kHz) resolution, and had an ionospheric activity metric value associated with them. In all cases, the most recently processed calibration was used, with processing dates ranging from 2017--2019. The same RTS version was used for all processing. Most data that satisfy the requirements are from 2013--2017. This search yielded 13,591 2-minute observations.

\subsection{Quality Assurance Metrics}
\begin{figure}
\includegraphics[width=0.5\textwidth]{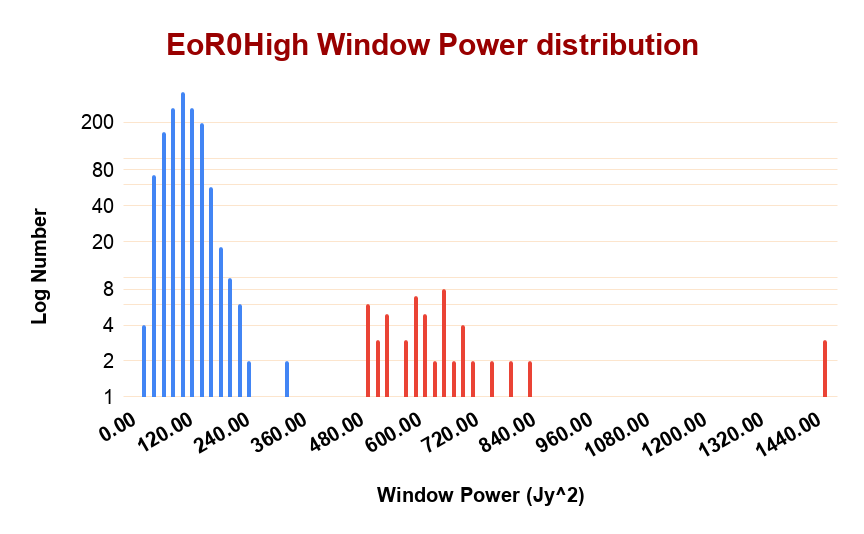}
\caption{Histogram of Window Power values for the EoR0 high-band observations, showing a clear multi-modal distribution. Blue data are included in further analysis, while red observations are omitted.}
\label{fig:distribution}
\end{figure}

There are four main quality assurance metrics applied to refine the dataset to be selected for power spectrum analysis: (1) calibration success (no errors; all frequencies present with finite-valued data), (2) ionospheric activity, (3) delay-space EoR Window Power, and (4) delay-space EoR Wedge Power. Here we describe these metrics and the order in which they are applied.
\begin{enumerate}
    \item Calibration: Data are calibrated using the MWA Real-Time System \citep[RTS, ][]{mitchell08,jacobs16,trott16}; the RTS performs direction-independent and direction-dependent (DD) calibration. It uses the primary beam model from \citet{sokolowski17} to select the 1000 apparent brightest sources for each snapshot pointing for calibration. The brightest five are used for DD calibration with a full Jones matrix solution for each source. These DD solutions are applied to the full set of 1000, effectively peeling five and directly subtracting the remaining 995. These calibrated data are fed through a validation process that confirms existence of all spectral channels with finite-valued data;
    \item Ionospheric Activity: we use the ionospheric activity metric developed by \citet{jordan17}, which uses the measured versus expected source positions of 1000 point sources in the field-of-view to estimate an ionospheric phase screen, and derive an activity metric that combines median source offset with source-offset anisotropy. Different thresholds are set for Phase I and Phase II datasets, where the reduced angular resolution of the latter array configuration yields a higher base activity level;
    \item Window Power: we use the delay transform power spectrum estimator to compute the power in the EoR Window, below the MWA's first coarse channel harmonic. Power is computed incoherently across baselines. Window Power is computed in the region bounded by the main beam lobe and the first coarse channel harmonic, for baselines with length $<100\lambda$ ($0<|\vec{u}|<100\lambda$, $k_\bot < k_\parallel < 0.4$);
    \item Wedge Power: we use the delay transform power spectrum estimator to compute the power below the EoR Window, in the primary beam main lobe wedge. Power is computed incoherently across baselines. The Wedge Power is computed for a region bounded by $k_\parallel=0$ and the main beam lobe wedge, also for baselines with length $<100\lambda$ ($0<|\vec{u}|<100\lambda$, $k_\bot > k_\parallel > 0$). Both are normalised by the number of contributing cells to yield an average power per cell (Jy$^2$).
\end{enumerate}
Using these metrics, cuts are made as follows, with the intent that unusually high-valued, or unusually low-valued snapshots are omitted. We take the full dataset and compute the average and standard deviation wedge and window power metrics for each data-type (Phase I or II, and high- and low-band). We omit snapshots that have wedge and window power values that fall outside of the primary mode of the distribution of values, and also snapshots that show consistent wedge power but high window power. The ionospheric cuts are made based on the metric values that are expected to produce biased results according to the analysis of \citet{trott17}. After the cuts are applied, datasets are then ordered by ionospheric metric value only. The distributions of Window Power are not Gaussian, and are generally multi-modal. Figure \ref{fig:distribution} displays the distribution for EoR0 high-band, with blue denoting included observations, and red denoting omitted. There is a clear separation of the two clusters, suggesting that calibration errors, and not statistical fluctuations, are the primary cause. This distinct behaviour is also observed for the other datasets. The data cuts are conservative insofar as all observations in the primary mode are included for all datasets.

\begin{figure}
\includegraphics[width=0.5\textwidth]{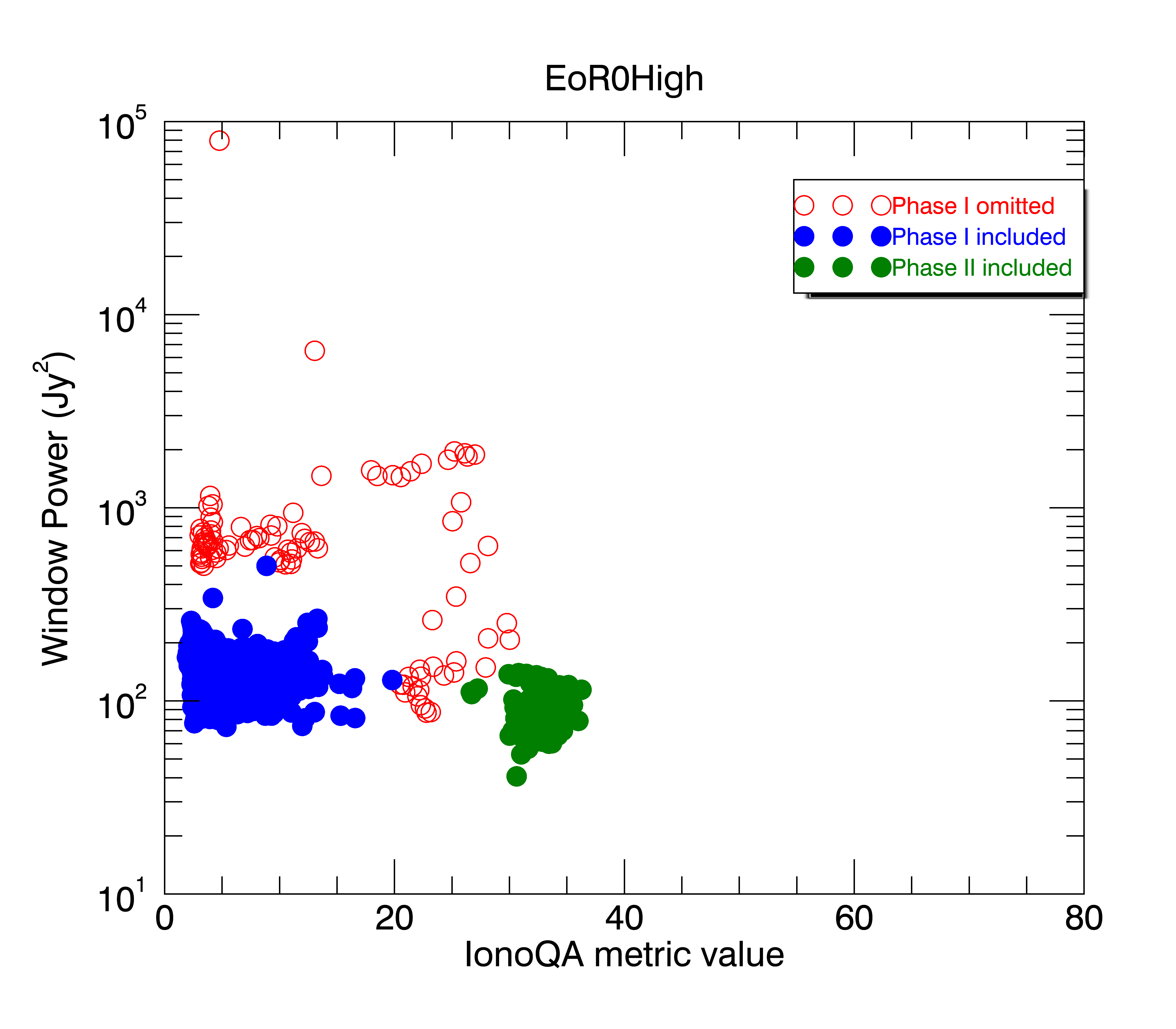}
\caption{Ionospheric activity metric versus power in the EoR window for the 1,581 observations in EoR0 high-band zenith data. The ionospheric metric captures anisotropy and turbulence produced by the ionosphere. Blue and green filled circles denote observations that meet the assessment criteria for Phase I and II, respectively. Open red circles are Phase I observations that are omitted due to high Window power or high ionospheric activity.}
\label{fig:eor0high_metrics}
\end{figure}

\begin{figure}
\includegraphics[width=0.5\textwidth]{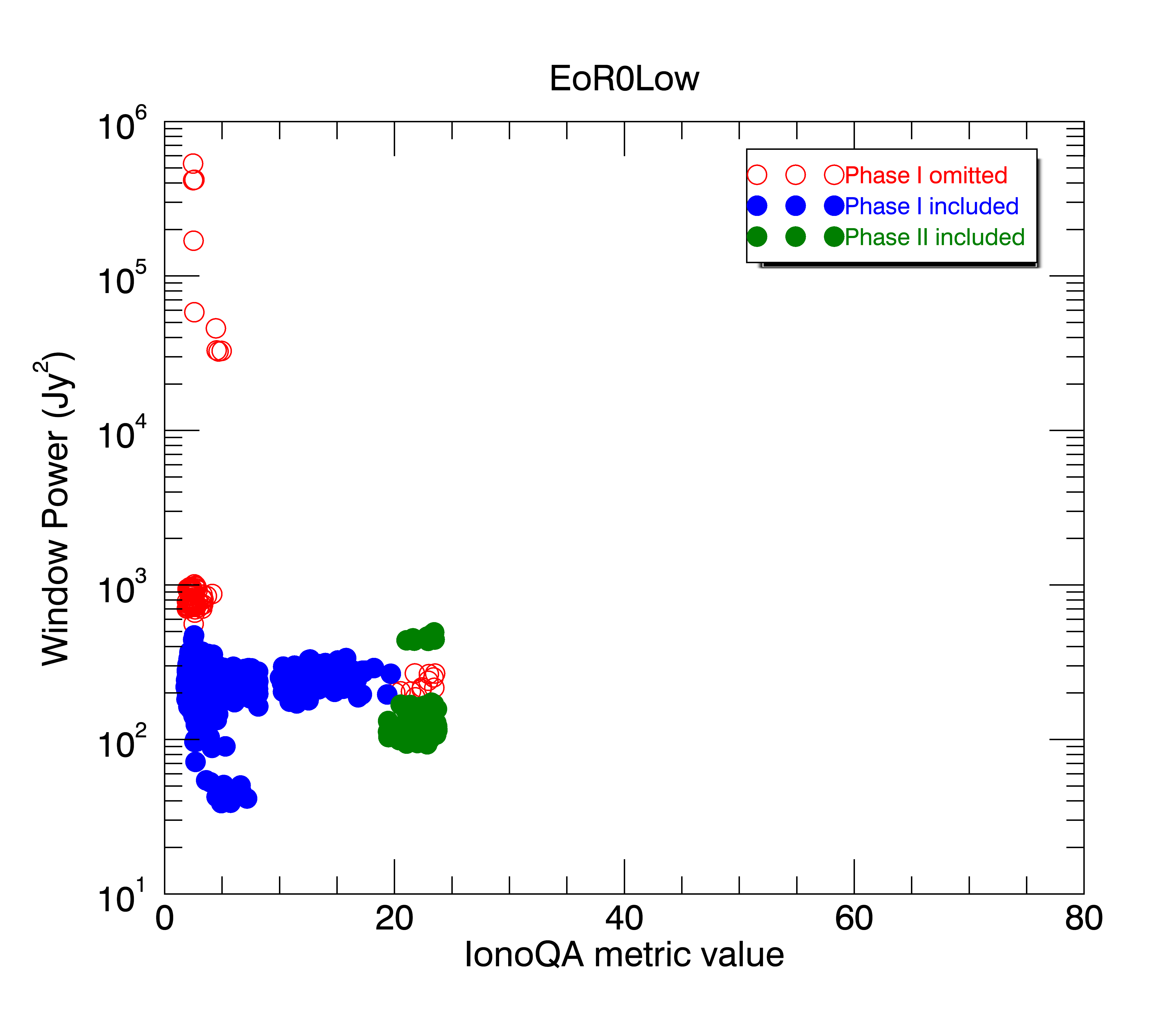}
\caption{Ionospheric activity metric versus power in the EoR window for the 1,761 observations in EoR0 low-band zenith data. Blue and green filled circles denote observations that meet the assessment criteria for Phase I and II, respectively. Open red circles are Phase I observations that are omitted due to high Window power or high ionospheric activity.}
\label{fig:eor0low_metrics}
\end{figure}

\begin{figure}
\includegraphics[width=0.5\textwidth]{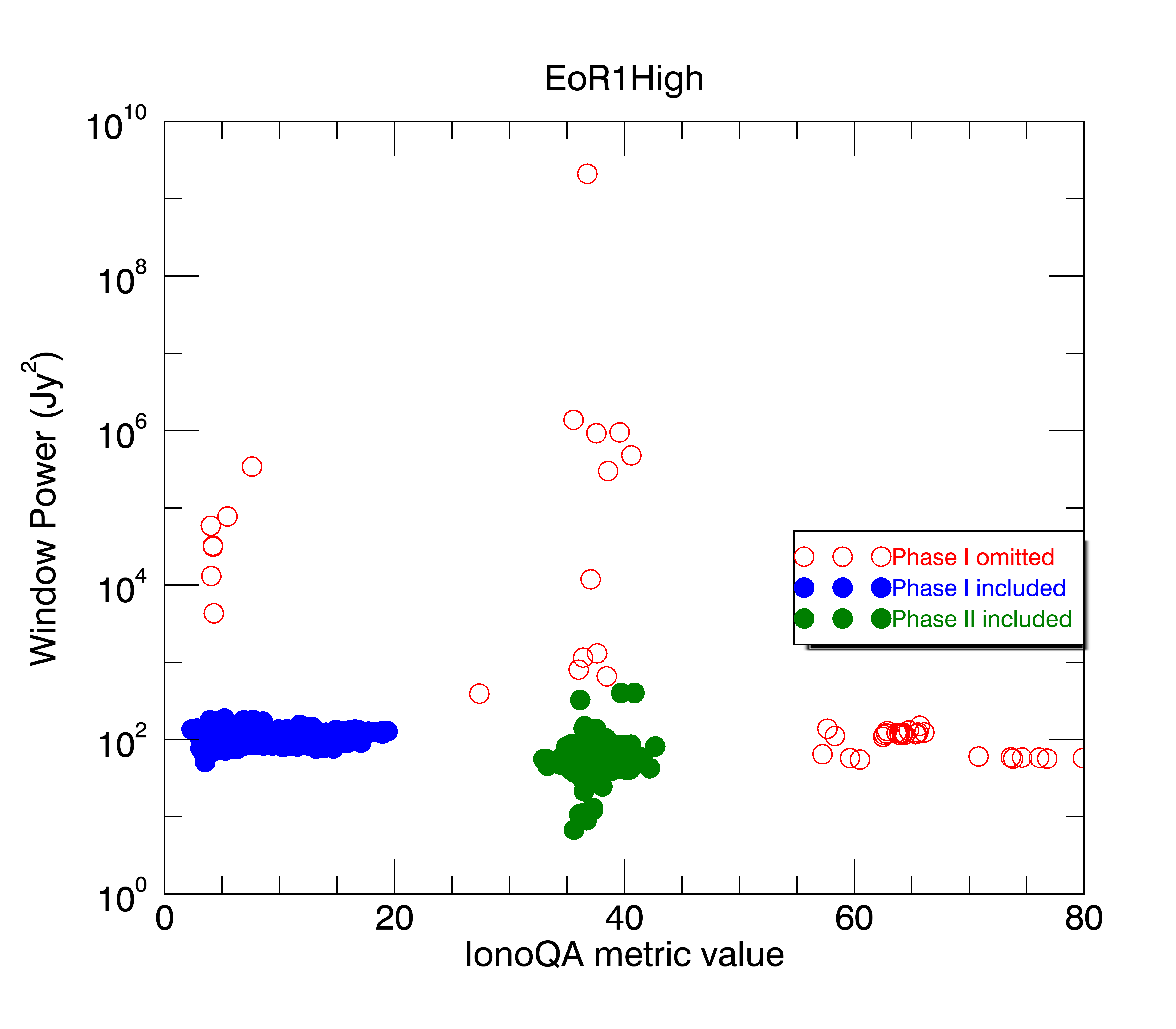}
\caption{Ionospheric activity metric versus power in the EoR window for the 1,123 observations in EoR1 high-band zenith data.}
\label{fig:eor1high_metrics}
\end{figure}

\begin{figure}
\includegraphics[width=0.5\textwidth]{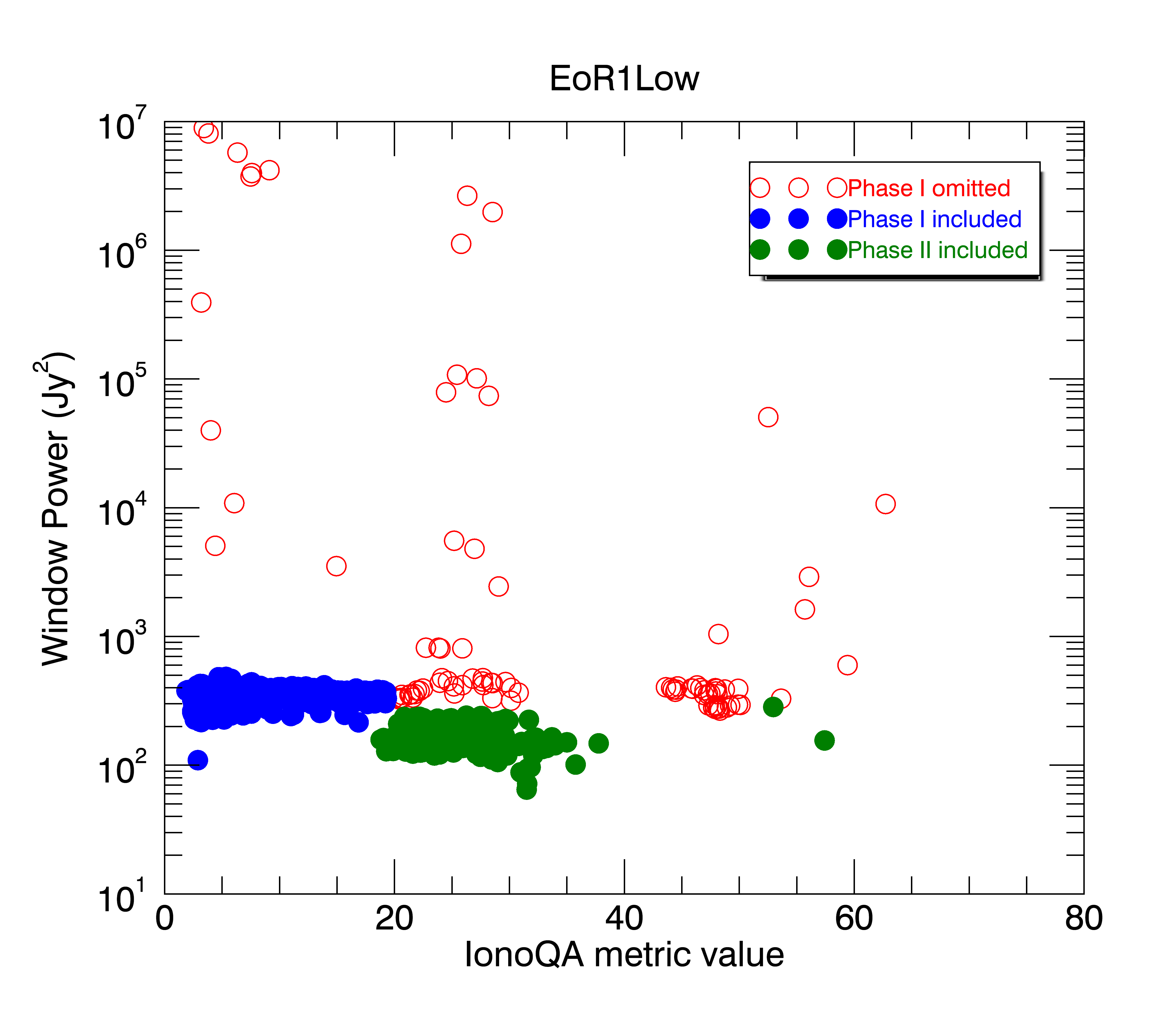}
\caption{Ionospheric activity metric versus power in the EoR window for the 1,814 observations in EoR1 low-band zenith data.}
\label{fig:eor1low_metrics}
\end{figure}

Table \ref{table:bispectra} describes the components of each dataset, and the changes after each stage of assessment.
\begin{table}{Datasets: total snapshot observations}
\centering
\begin{tabular}{|c||c|c|c||c|}
\hline 
Field & Total & IonoQA & Final & Cuts* \\
\hline
EoR0High & 4187 & 4108 & 3890 & P$_{\rm max}$=300,Iono$_{\rm max}$=8 (50)\\
EoR1High & 1123 & 1084 & 986 & P$_{\rm max}$=300,Iono$_{\rm max}$=30 (50)\\
EoR2High & 1646 & 1646 & 1575 & P$_{\rm max}$=300,Iono$_{\rm max}$=30 (50)\\
EoR0Low & 3252 & 3104 & 2901 & P$_{\rm max}$=300,Iono$_{\rm max}$=8 (50)\\
EoR1Low & 1814 & 1806 & 1708 & P$_{\rm max}$=500,Iono$_{\rm max}$=30 (50)\\
EoR2Low & 1569 & 1569 & 1509 & P$_{\rm max}$=300,Iono$_{\rm max}$=8 (50)\\
\hline
Total & 13,591 & 13,317 & 12,569 & \\
\end{tabular}
\caption{Datasets used in this analysis, including the original extracted sets of observations ('Total'), those remaining after calibration and ionospheric cuts (`IonoQA'), and the final datasets after Window and Wedge power cuts (`Final'). The `Cuts' column describes the ionospheric metric and Window Power thresholds set for the final datasets for Phase I data. *Window Power thresholds for Phase II are a factor of two lower, due to the larger number of baselines contributing on these scales.}\label{table:bispectra}
\end{table}
Figures \ref{fig:eor0high_metrics} -- \ref{fig:eor1low_metrics} show examples of power in the EoR window versus ionospheric activity for different fields and frequencies. The omitted observations plotted in these figures are all from Phase I; Phase II observations extracted from the database were consistent with a quiet ionosphere.
Blue and green filled circles denote observations that meet the assessment criteria for Phase I and II, respectively. Open red circles are observations that are omitted due to high Window power or high ionospheric activity. These cuts remove 10--20 percent of the data. The majority of the removed data contain particular tiles or baselines with very poor calibration, leading to excess power in those modes.

\subsection{Simulations}\label{sec:simulations}
Correctly calculating the normalisation from measurement units (Jansky and Hertz) to cosmological units (megaparsecs) seems trivial, but is complicated by the choices made during the analysis pipeline (e.g., cross-multiplying even and odd samples for the cross power spectrum, and the definition of Stokes I with respect to polarisation axes). It is important to ensure that the normalisation is correct, and that signal is not being lost due to coherence of coherently-gridded data. The former can be ensured via matching of different approaches to calculating the noise, and to internal consistency between independent MWA analysis pipelines \citep{barry19}. The latter can be achieved using a signal simulation; ensuring that the input power spectrum is recovered after passing through the CHIPS pipeline.

We performed a large simulation using a modified version of 21cmFAST \citep{mesinger11,greig20}, tuned to the MWA's large primary beam size and frequency resolution. A box with 6,400 pixels on each side was produced, with 7,500 cMpc on angular sides, and 1.172 cMpc line-of-sight resolution. This large simulation is important for creating simulations that can be directly applied to MWA, without need for padding or interpolation. The data are produced as a light-cone, with signal evolution as a function of redshift. We apply a beam model to the slices, perform an angular Fourier Transform, and convert the brightness temperature units to Jansky per steradian.

\begin{figure}
\includegraphics[width=0.5\textwidth]{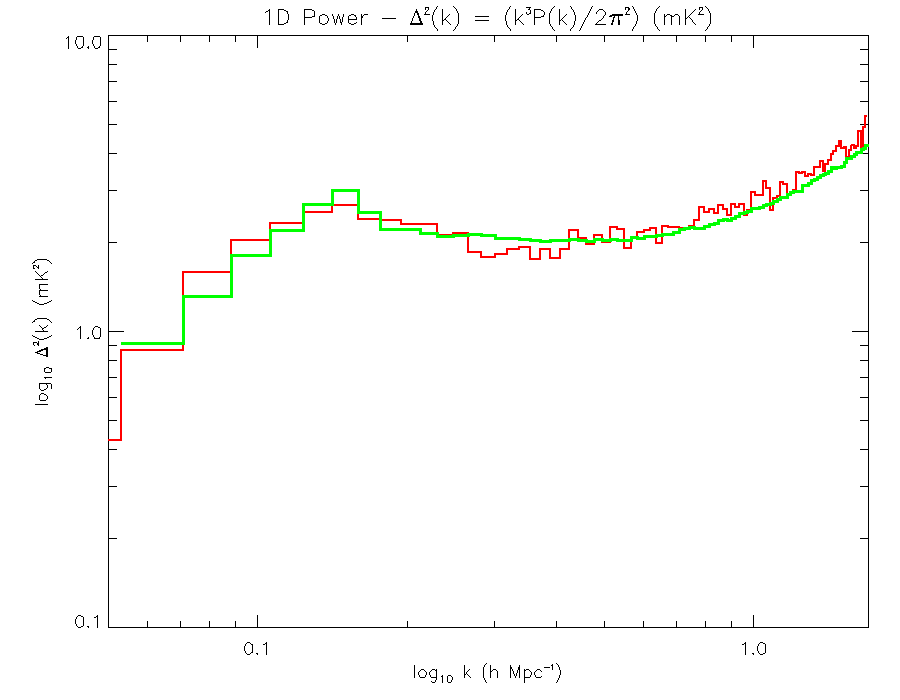}
\caption{Simulated EoR signal from 384 channels at $z$=6.8; input (green) and recovered through CHIPS pipeline (red), using a simulation tuned to the MWA observational parameters, and underpinned by a 21cmFAST model for the EoR signal. This demonstrates that signal power is not being lost through decoherence in the analysis, with consistency in both the shape and amplitude.}
\label{fig:simulation}
\end{figure}

Figure \ref{fig:simulation} shows the input EoR 21cm power spectrum over 384 channels centred at $z=6.8$ (green) and the signal power recovered through CHIPS (red). This demonstrates good consistency of the input and output signal levels in both shape and amplitude.
Given that we do not perform any post-calibration subtraction of signal from the data, these results provide confidence that signal loss is not occurring in the CHIPS pipeline.

\begin{figure}
\includegraphics[width=0.5\textwidth]{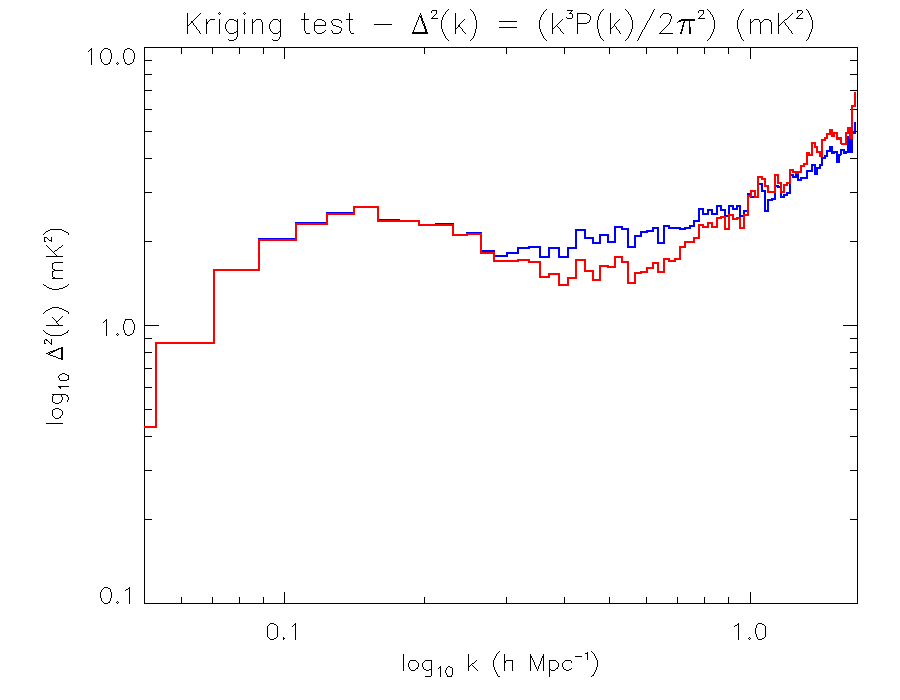}
\caption{Simulated EoR signal from 384 channels at $z$=6.8 for the full dataset (red) and a dataset with the MWA's missing channels, where kriging in-painting has been applied (blue). For modes in the EoR Window ($k < 0.4h$Mpc$^{-1}$), there is no bias caused by the kriging.}
\label{fig:kriging}
\end{figure}

In addition to the regular simulation, we also perform the same operation but with the missing channels corresponding to those in the actual data. We apply the kriging to the mitigate the missing channels and check that the output power levels are still consistent. This procedure demonstrates that (1) the kriging is not biasing the results, and (2) that kriging is offering no benefit at small $k$, but does close to the coarse channel harmonics. Note that we do not use those harmonic modes in our measurements. Figure \ref{fig:kriging} shows the 1D power spectra from the same simulation with the kriging applied (blue) and the full dataset (red). The contaminated modes, where there is a discrepancy, correspond to the location of the coarse channel harmonic. The primary results shown in this work are for $k < 0.4~h$Mpc$^{-1}$.

We also compare simulation outputs through RTS-CHIPS with those through FHD-$\epsilon$ppsilon \citep{barry19a}, to ensure power and noise-level consistency. A similar EoR simulation was produced and passed through both pipelines, yielding consistent results (N. Barry, in prep.). Although this deviates somewhat from previous MWA EoR papers where the actual data sets were processed through both pipelines \citep[Figure 7,]{barry19a,li19, beardsley16}, this retains the same philosophy of ensuring robustness of results using independent calibration and analysis methods.

\section{Results}\label{sec:results}
The data are divided into their respective fields and observing bands. In order to present results that are cosmologically-relevant, the data need to be analysed in sub-bands to ensure signal ergodicity within the volume (i.e., the signal does not evolve over the bandwidth of the experiment). For these data, we use a set 15.36~MHz band, which, after tapering by the line-of-sight Blackman-Harris window function, yields an effective bandwidth of $\sim$10~MHz. For our purposes, this amounts to three overlapping 15.36~MHz bands (192 channels) within the 30.72~MHz (384 channels), with a 96-channel separation between the centres of each. This leads to a correlation level of 3\% within each band between redshift bins, and 0\% between bands, which should be accounted for in future parameter estimation work.

In addition to the different fields and bands, some datasets have both zenith and off-zenith pointings. Ideally, given that these observation sets have the same phase centre, these should be able to be added coherently to obtain the best thermal noise reduction. However, the instrument response changes between pointings, and it is important that we can demonstrate that coherent addition does not lead to signal loss through decoherence.

\begin{figure}
\includegraphics[width=0.5\textwidth]{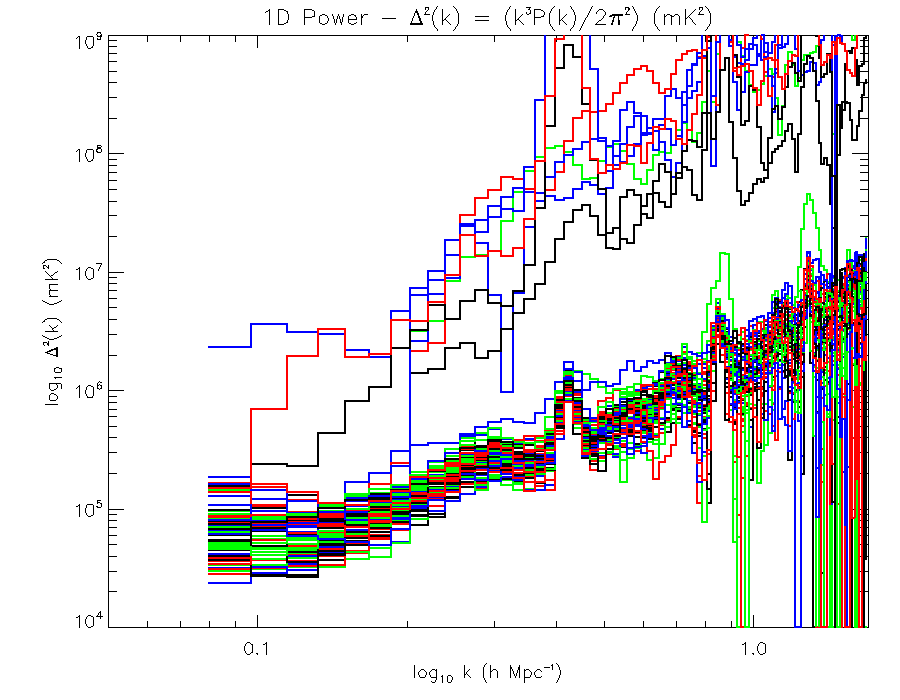}
\caption{All sets of 20 observations from the EoR0Low zenith-pointed dataset. Most sets have results that are clustered at low power, but some contain contaminated modes that spread high power to all scales.}
\label{fig:eor0low_sets}
\end{figure}

For all data, we take consistent cuts to average from 2D to 1D: $k_\parallel > 3.5 k_\bot$; $k_\parallel > 0.15 h$Mpc$^{-1}$; $0.01 \geq k_\bot \geq 0.04h$Mpc$^{-1}$. These cuts are bounded by the angular modes that are well-sampled by the MWA baselines, and line-of-sight modes that lie outside of the foreground horizon line (plus a $k=0.05h$Mpc$^{-1}$ buffer). Note that formation of 1D power spectra is made directly from the 3D modes, and not through 2D. We start by displaying some of the range of data found in each of the sets of 20 observations, as indicative of the qualitative difference between clean and contaminated data. Figure \ref{fig:eor0low_sets} shows the 1D full-band power spectra from the EoR0Low field zenith sets, as an example.
There is a clear distinction between the clustered data and the erroneous data. These contaminated datasets may only contain one bad observation, but that can be enough to cause excess power. We subsequently remove the clearly-contaminated data, and retain only the clustered data. We are careful to cut conservatively, in order to avoid biasing the data by cleaning normal, but statistically high, data. Some datasets show no contaminated sets, and all data are retained for further analysis. Despite the Window and Wedge power cuts made to the initial data, outlier power spectra can still arise. This is due to the power spectra being formed from a smaller subset of the data than used for the cuts, and the conservative cuts made initially to the datasets.

In the results that follow, kriging to in-paint the missing channels is always applied with the same set of hyperparameters. In some cases, this does a poor job to cleanly smooth over the missing data, most notably for the low-band data. Instead of trying to optimise the hyperparameters and potentially biasing the results, we leave them as fixed and accept poorer performance in these modes. The source of the poor performance is not likely the missing channels themselves, but an increased signal variance observed in edge channels for some datasets that is due to poor bandpass calibration.

\subsection{Individual sets - spherically-averaged power spectra}
We present a census of the data from four years of the MWA EoR experiment; the cleanest subsets of data assessed in this work, taken from three observing fields, two broad observing spectral bands, and multiple telescope pointings (for EoR0 and EoR2).

\subsubsection{E0R0}
\begin{figure*}
\includegraphics[width=0.32\textwidth,angle=0]{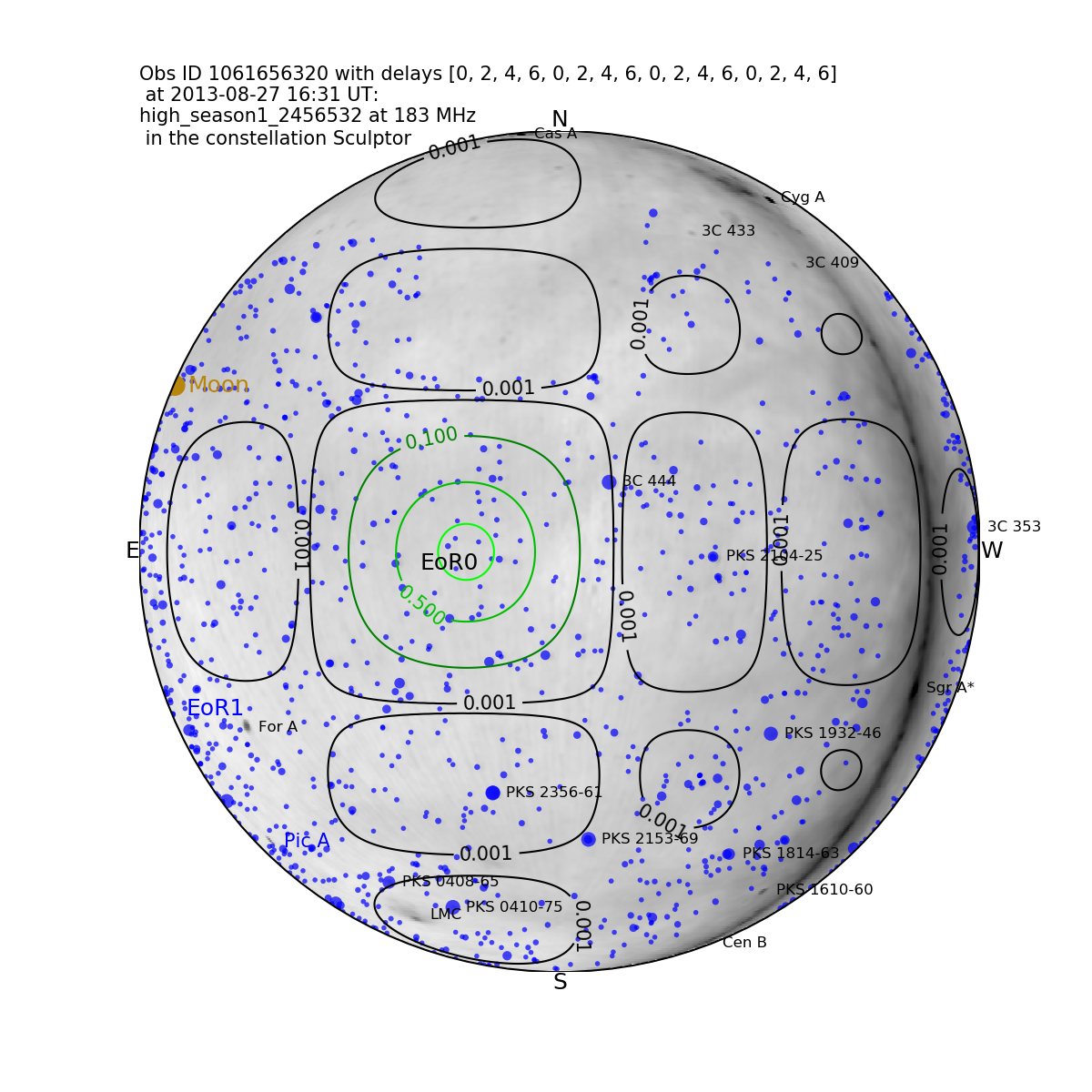}
\includegraphics[width=0.32\textwidth,angle=0]{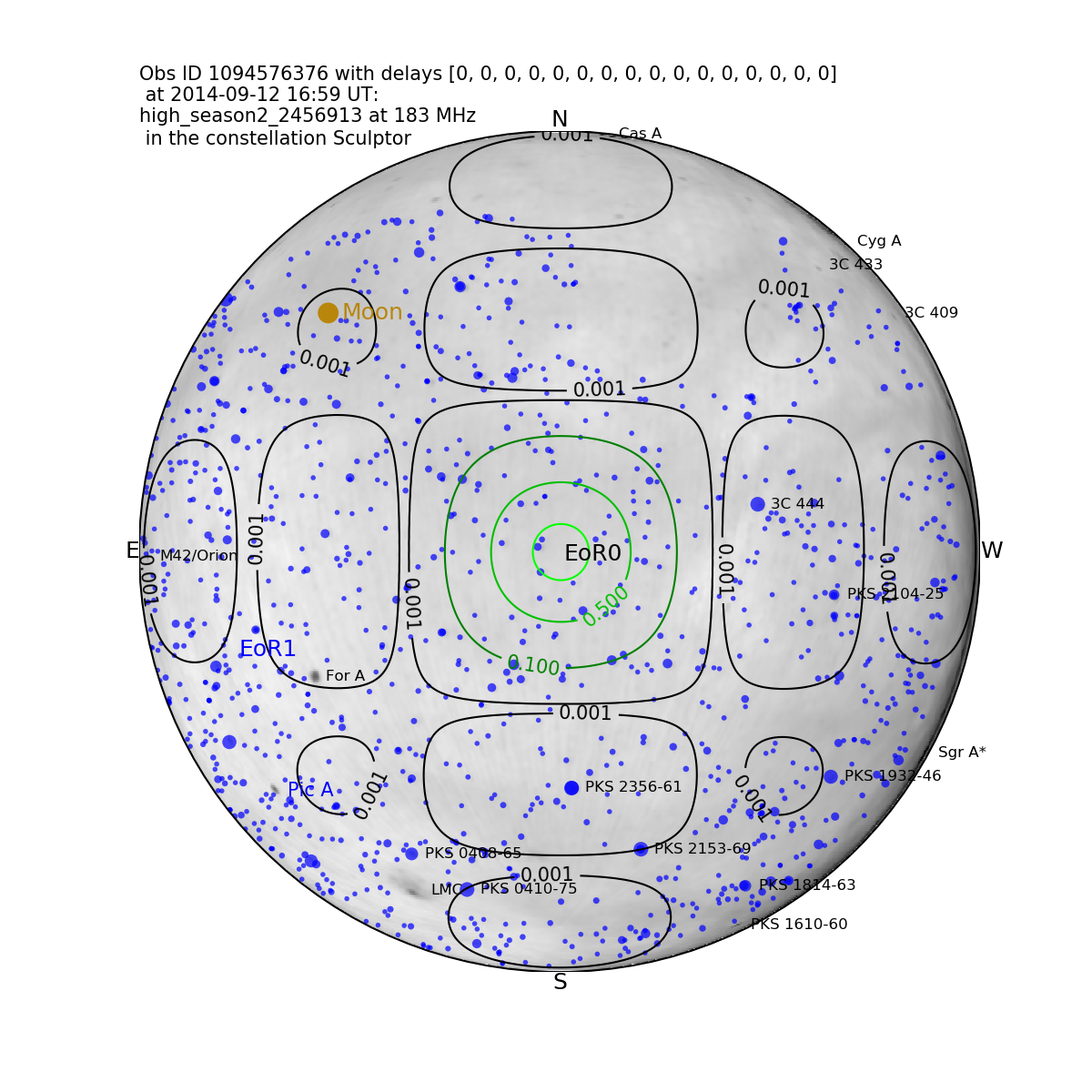}
\includegraphics[width=0.32\textwidth,angle=0]{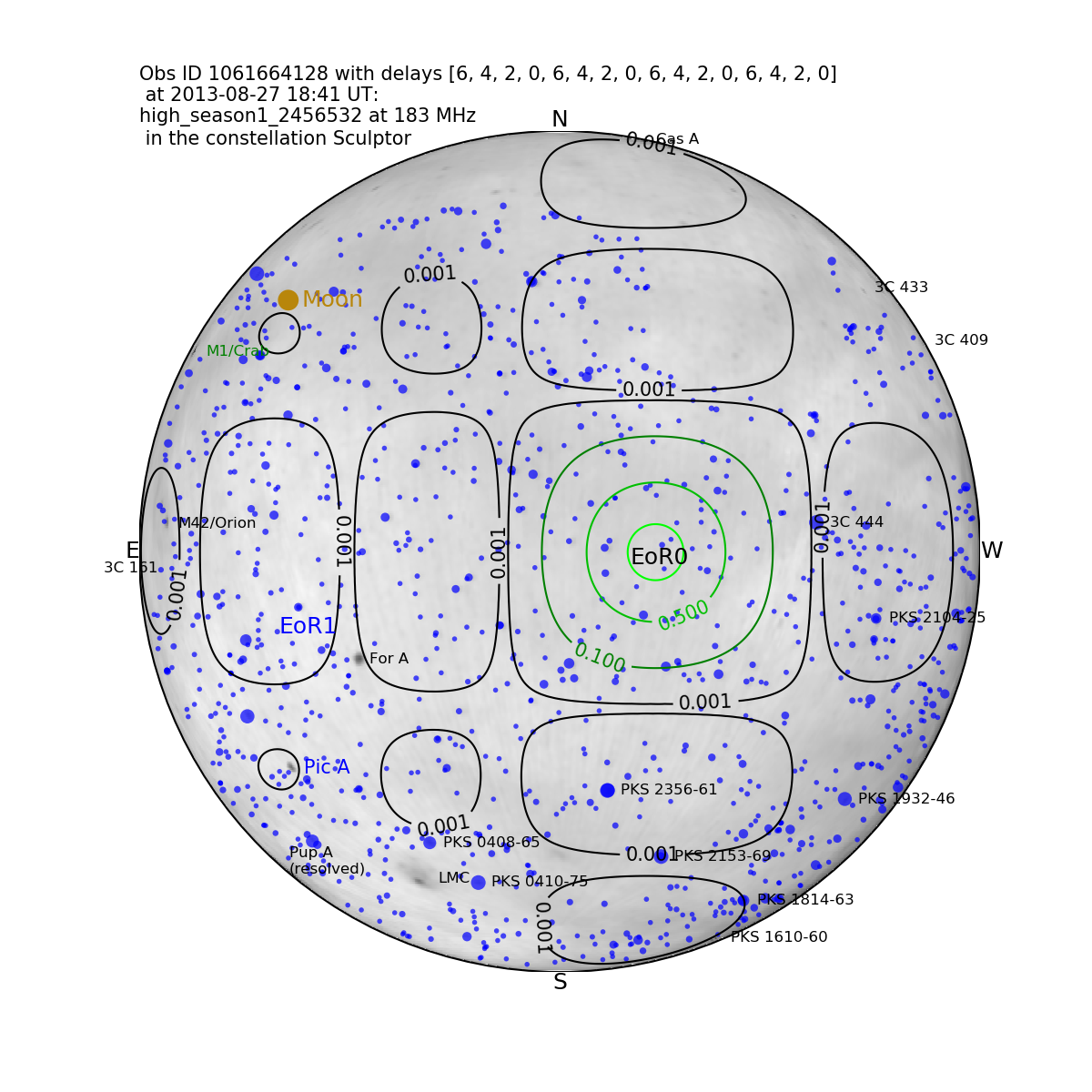}
\caption{Sky response of the telescope at the central frequency for three pointings of the EoR0 field in the high band: Minus2 (left), Zenith (centre), Plus2 (right). The Galactic Centre is setting and prominent in the Minus2 pointing. While the Galactic Centre is past the second sidelobe, it still affects the power spectrum.}
\label{fig:eor0high_sky}
\end{figure*}

\begin{figure*}
\includegraphics[width=1.\textwidth,angle=180]{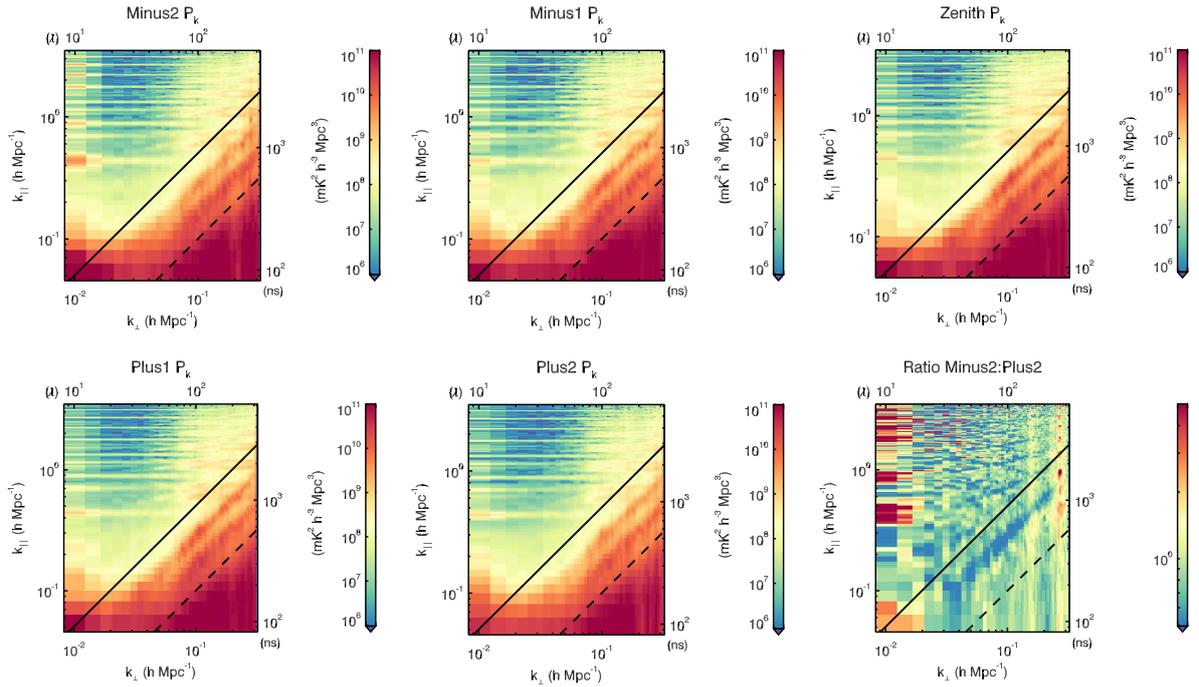}
\caption{Set of power spectra from different pointings for the EoR0 field and full high-band. The bottom-right plot shows the ratio of Minus2 to Plus2 pointings. The rotation of foreground structures through different primary beam responses combined with the evolution of the beam response with frequency, lead to shifting foreground contamination in parameter space. In particular, the ratio demonstrates the changing horizon power over pointings. The black solid (dashed) lines correspond to the horizon (first beam null) for a flat-spectrum point source.}
\label{fig:eor0high_5pointings}
\end{figure*}

EoR0 is centred at RA=0h, Dec=$-$27$^o$ and contains no major extended radio sources. It is the best-studied of the MWA EoR fields, with all currently-published results derived from it. Figure \ref{fig:eor0high_sky} shows the primary beam response at 180~MHz of the Minus2, Zenith and Plus2 pointings. The setting Galactic Centre is prominent in the Minus2 pointing. Figure \ref{fig:eor0high_5pointings} shows the 2D power spectrum for the best observations from the five central pointings (color-scale and parameter space have been reduced to highlight differences), and the ratio of Minus2 to Plus2 pointings.
The diagonal stripes of increased power show the sidelobes of the primary beam response, and the ratio shows the changing horizon power.

\begin{figure}
\includegraphics[width=0.5\textwidth,angle=0]{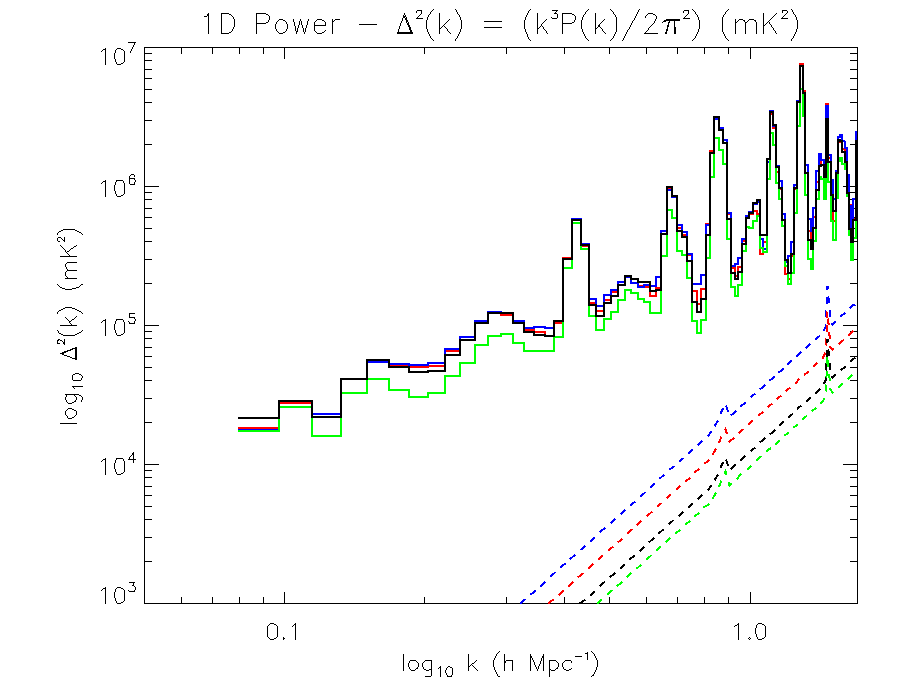}
\caption{Set of power spectra from the zenith pointing of EoR0 high-band for increasingly-larger subsets of the full set: 300 (blue), 500 (red), 680 (black), 820 (green) observations. In this plot, and all subsequent 1D plots, the green diagonal lines denote 2$\sigma$ thermal noise. Colours are matched between the power and the diagonal noise curves (dashed).}
\label{fig:adddata}
\end{figure}

EoR0 high-band zenith contains the largest number of observations in the whole dataset. To test the utility of continuing to add further zenith data, and the usefulness of the metrics we have used for data selection, we can study the 1D power spectrum for different subsets of the data: 300, 500, 680, 820 observations (Figure \ref{fig:adddata}).
There is little improvement in adding extra data from 300 observations onwards, until the final aggregation with 820 observations. The data are ordered only by ionospheric metric, and therefore the window power can change from observation-to-observation (after initial cuts). However, these data are all selected to be ionospherically-quiet and these results indicate that contamination in the EoR Window may be a stronger selector for high-quality data than ionospheric activity. To test this, we order the EoR0 high-band data by window power, and compute the 1D power spectra for the first 20 sets of 20 snapshots. This results in power levels in the $k=0.07-0.2h$Mpc$^{-1}$ range that are 1.5--1.7 times lower than ordering on ionospheric metric, consistent with the window power being a stronger selector. In either case, the power spectra are clearly systematics-dominated, exceeding the thermal noise level across most scales.

\begin{figure}
\includegraphics[width=0.5\textwidth]{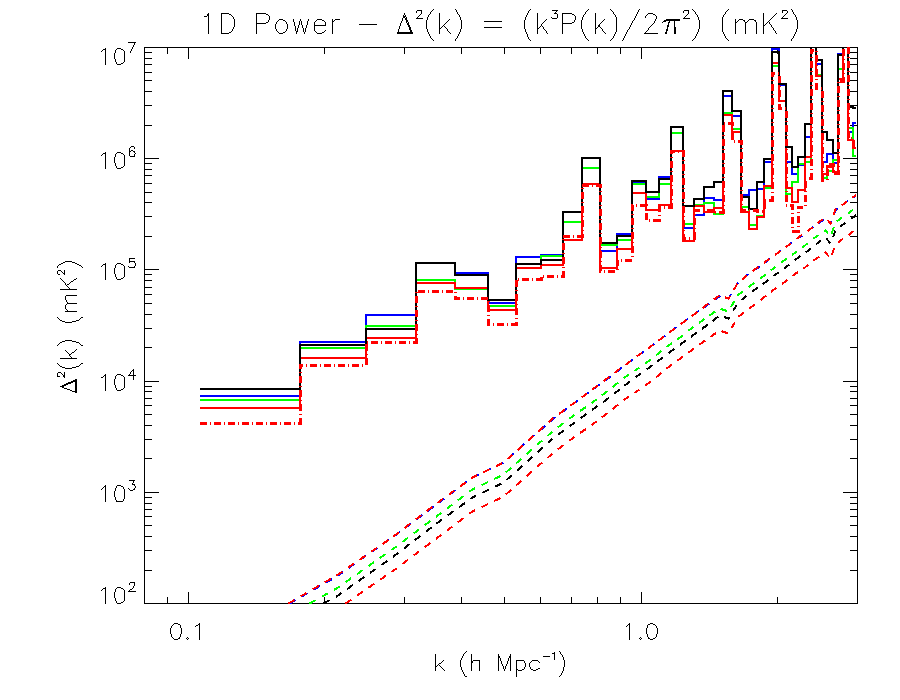}
\caption{Measured 1D power spectrum for the best 3,340 observations (aggregated, 111 hours) from EoR0 high-band across five pointings at $z=6.5$ (red=zenith, black=Minus2, blue=Minus1, green=Plus1, red-dashed=Plus2). Colours are matched between the power and the diagonal noise curves (dashed).}
\label{fig:1d_eor0high_192chan_3340_all}
\end{figure}

Figure \ref{fig:1d_eor0high_192chan_3340_all} shows the measured 1D power spectra from the five central pointings and the lowest redshift, $z=6.5$ (182--197~MHz). They are broadly consistent. Thermal noise curves at 2$\sigma$ are shown as the green diagonal lines for the zenith (1000 observations) and Minus1 pointing (500 observations) to give indicative noise levels.
The Galactic Centre having set in the later pointings (Plus1, Plus2) seems to translate into lower power for these.

\begin{figure}
\includegraphics[width=0.5\textwidth]{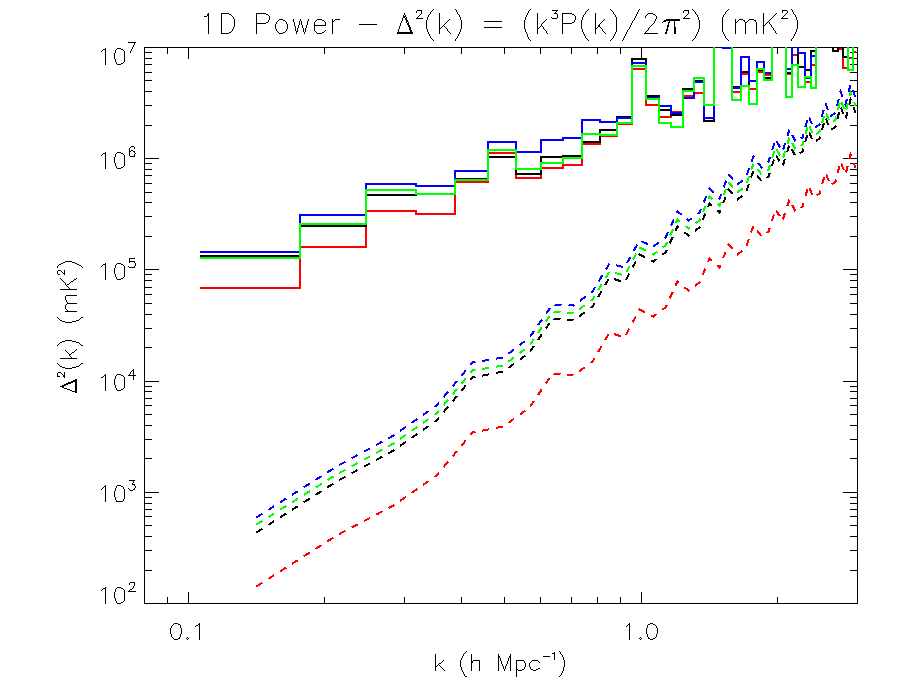}
\caption{Measured 1D power spectrum for the best 1,140 observations (aggregated, 38 hours) from EoR0 low-band across four pointings at $z=7.8$ (red=zenith, black=Minus2, blue=Minus1, green=Plus1). Colours are matched between the power and the diagonal noise curves (dashed).}
\label{fig:1d_eor0low_192chan_1140_all}
\end{figure}

Figure \ref{fig:1d_eor0low_192chan_1140_all} shows the equivalent lowest redshift band for the EoR0Low data ($z=7.8$).
The Plus2 pointing has been omitted due to the small dataset that remained after the quality assurance cuts were applied.
The results are poorer at the higher redshift. This is due to a combination of higher system temperature, increased ionospheric distortion, broader primary beam shape, and larger distance to the field, with the increased beam size capturing more Galactic emission of prime importance.

\subsubsection{E0R1}
\begin{figure*}
\includegraphics[width=0.45\textwidth,angle=0]{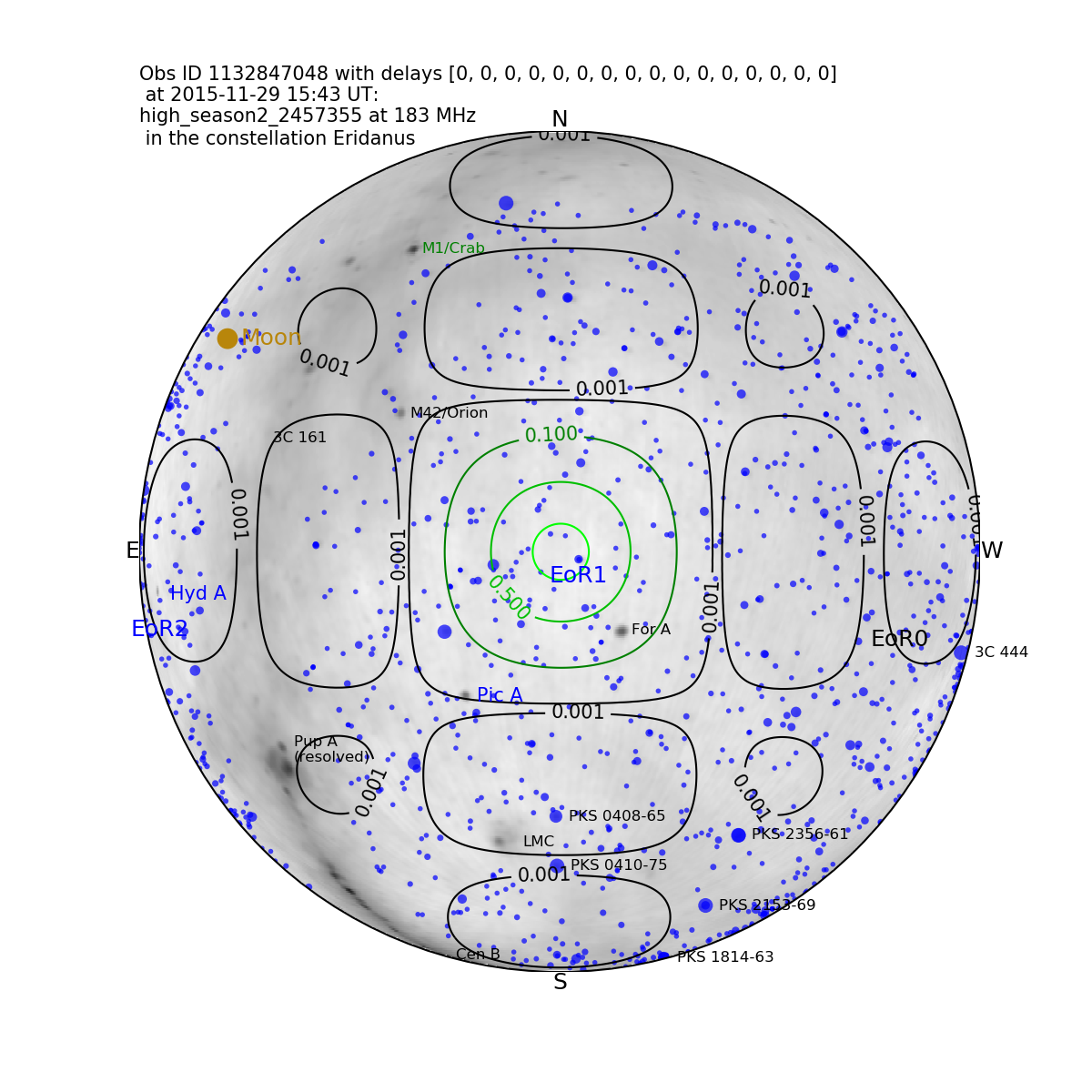}
\includegraphics[width=0.45\textwidth,angle=0]{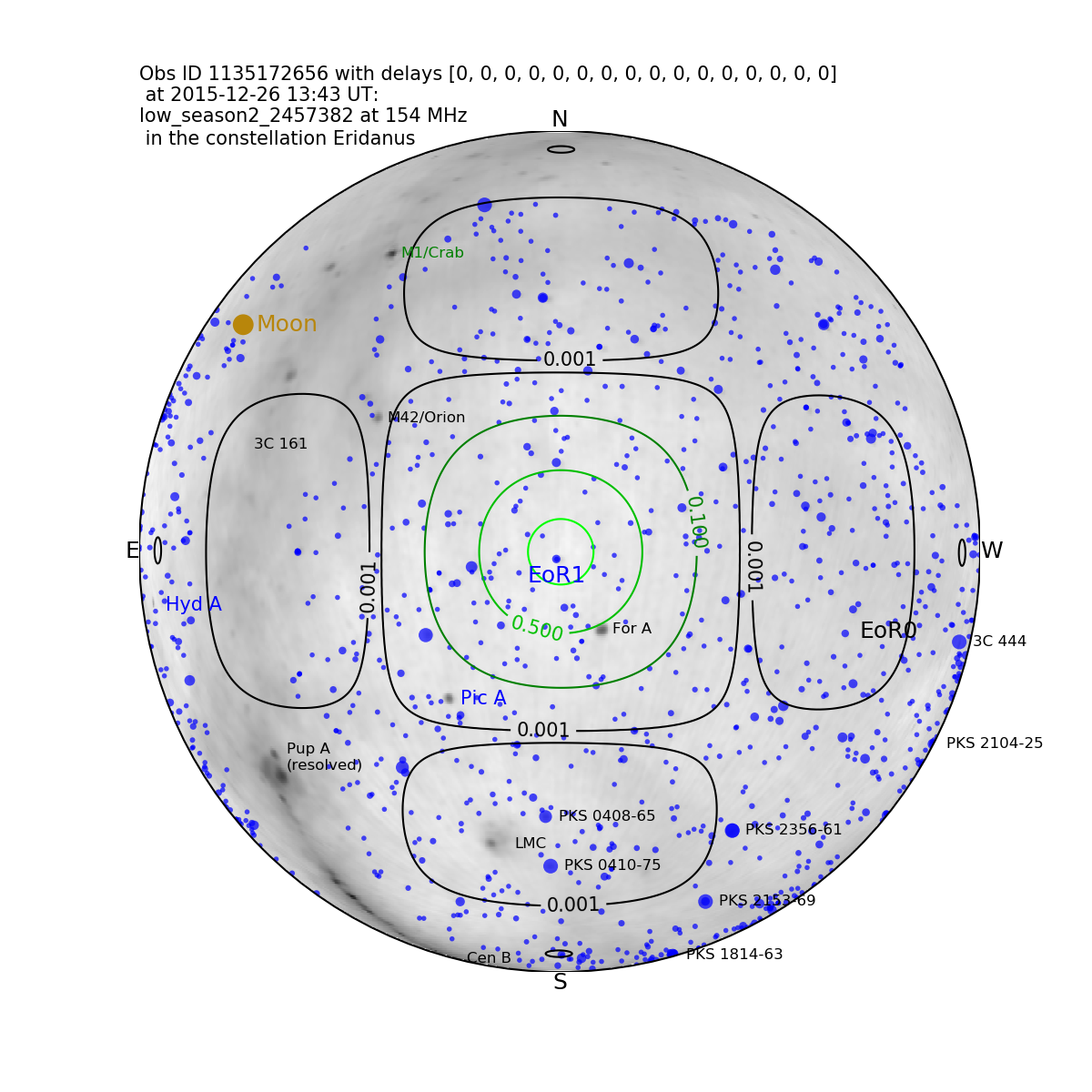}
\caption{Sky map (Haslam) overlaid with the MWA tile primary beam response contours at 180~MHz showing the response of the telescope for the high- and low-band of the EoR1. Fornax A is bright and extended, residing in a spectrally-complex part of the primary beam.}
\label{fig:eor1high_sky}
\end{figure*}

Figure \ref{fig:eor1high_sky} shows the sky map and primary beam response in the high- and low-band for the zenith pointing of the EoR1 field. Fornax A, a several hundred Jansky extended radio galaxy, appears in the main beam lobe, contributing tens of Janskys of apparent flux density into the data.
Previous work has shown that accurate removal of Fornax A is crucial for scientifically-useful results from the EoR1 field \citep{procopio17}. For these data, the Fornax A calibration model used a preliminary shapelet-based model. Future calibrations will use an improved shapelet model, which has been demonstrated to be more accurate than the previous shapelet model and a point source-based model \citep{line19}, although this does not seem to be the largest systematic in this field. We expect that results from this field will improve with the new calibration.

\begin{figure}
\includegraphics[width=0.5\textwidth]{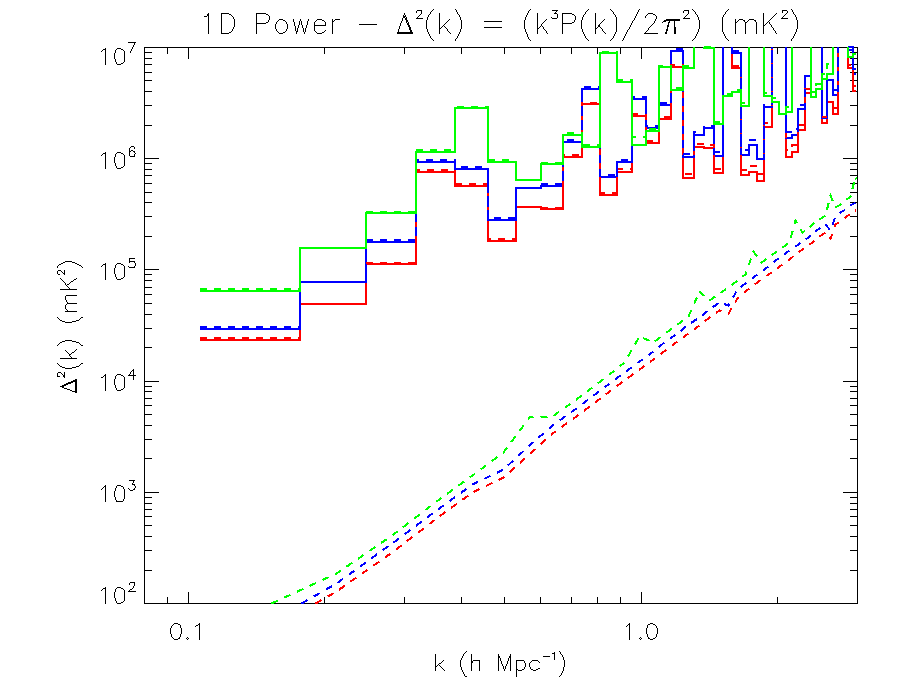}
\caption{Measured 1D power spectrum (solid line) and power plus 2$\sigma$ uncertainty (dashed) for the best 600 observations (20 hours) from EoR1 high-band from the zenith pointing at $z=6.5$ (red), $z=6.8$ (blue), $z=7.1$ (green).}
\label{fig:1d_eor1high_192chan_600}
\end{figure}

\begin{figure}
\includegraphics[width=0.5\textwidth]{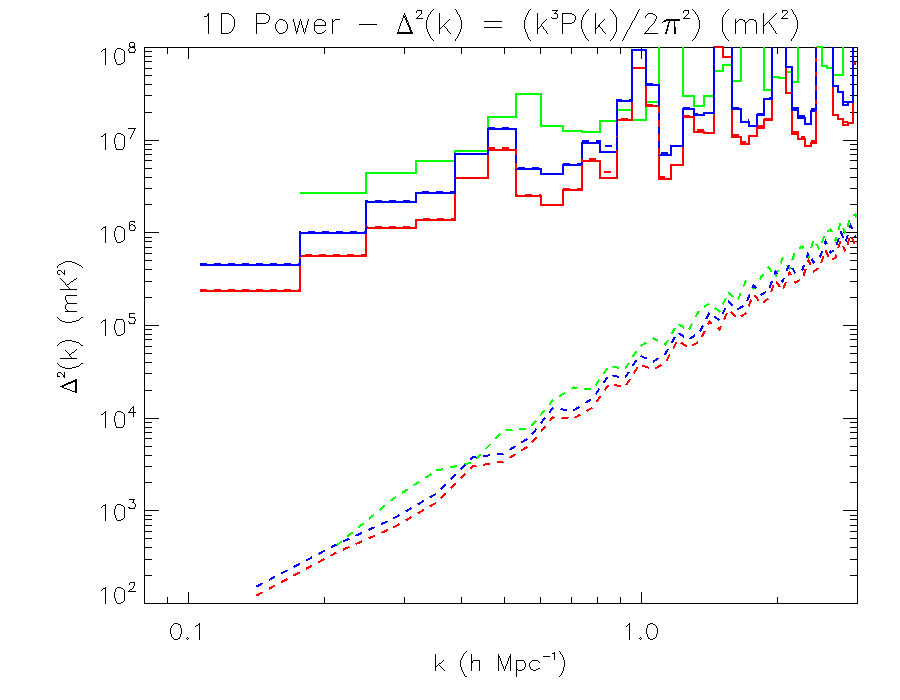}
\caption{Measured 1D power spectrum (solid line) and power plus 2$\sigma$ uncertainty (dashed) for the best 800 observations (27 hours) from EoR1 low-band from the zenith pointing at $z=7.8$ (red), $z=8.2$ (blue), $z=8.7$ (green).}
\label{fig:1d_eor1low_192chan_800}
\end{figure}

Figure \ref{fig:1d_eor1high_192chan_600} shows the measured 1D power spectrum (solid line) and the power plus 2$\sigma$ noise uncertainty for the best 600 observations (20 hours) from EoR1 high-band from the zenith pointing.
Figure \ref{fig:1d_eor1low_192chan_800} then shows the measured 1D power spectrum and power plus 2$\sigma$ noise uncertainty for the best 800 observations (27 hours) from EoR1 low-band from the zenith pointing.
The results are substantially poorer than those from the EoR0 field, likely owing to the more complicated, and less developed, sky model required to calibrate data and peel foregrounds.

The robustness of this conclusion, and the reproducibility of the increased power, can be tested by splitting the data into two equal datasets of 300 observations and computing the power. This produces power spectra that are statistically equivalent, suggesting that overall data quality for the EoR1 field with this calibration model is reduced, rather than particular poor observations contaminating the results.

\subsubsection{E0R2}
\begin{figure*}
\includegraphics[width=0.32\textwidth,angle=0]{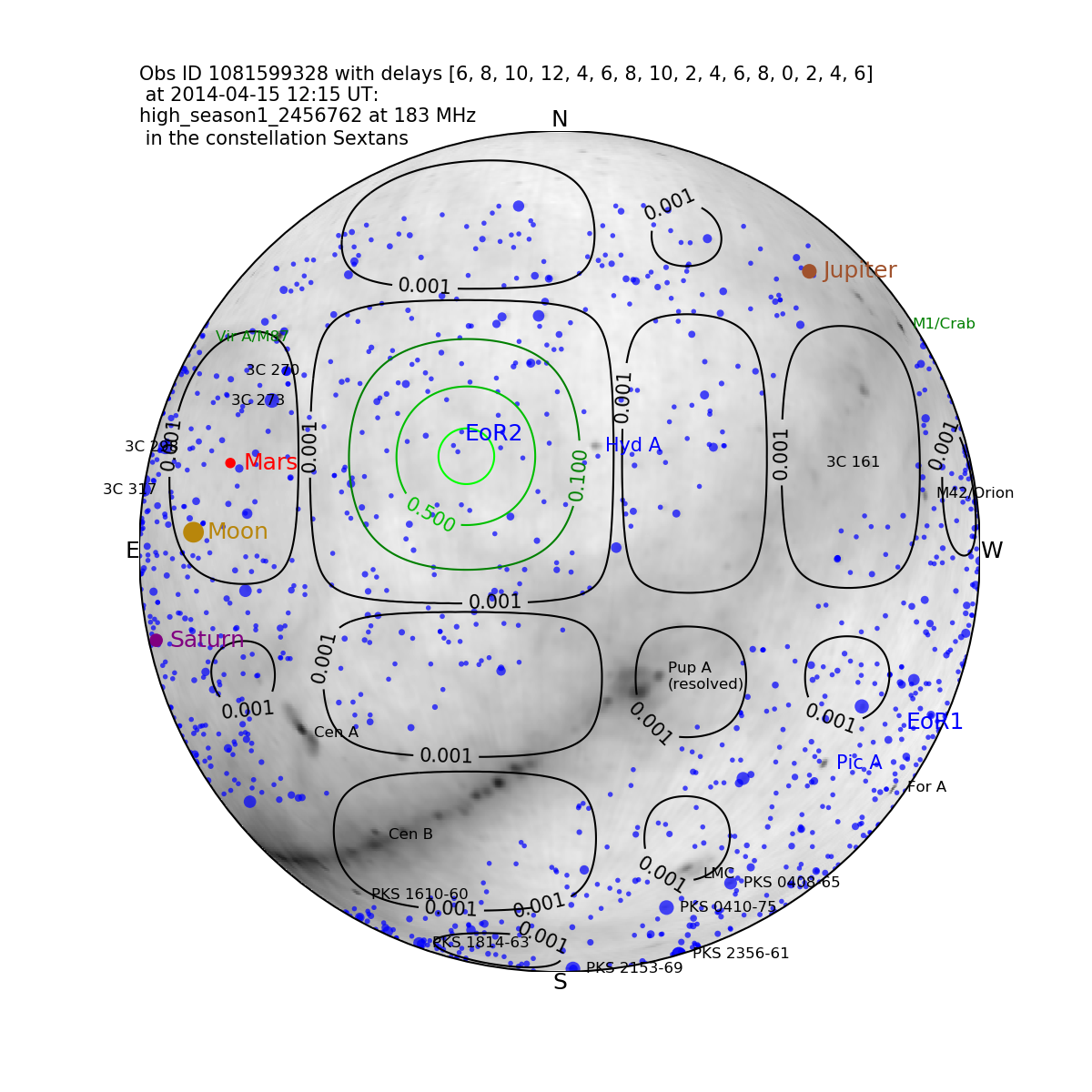}
\includegraphics[width=0.32\textwidth,angle=0]{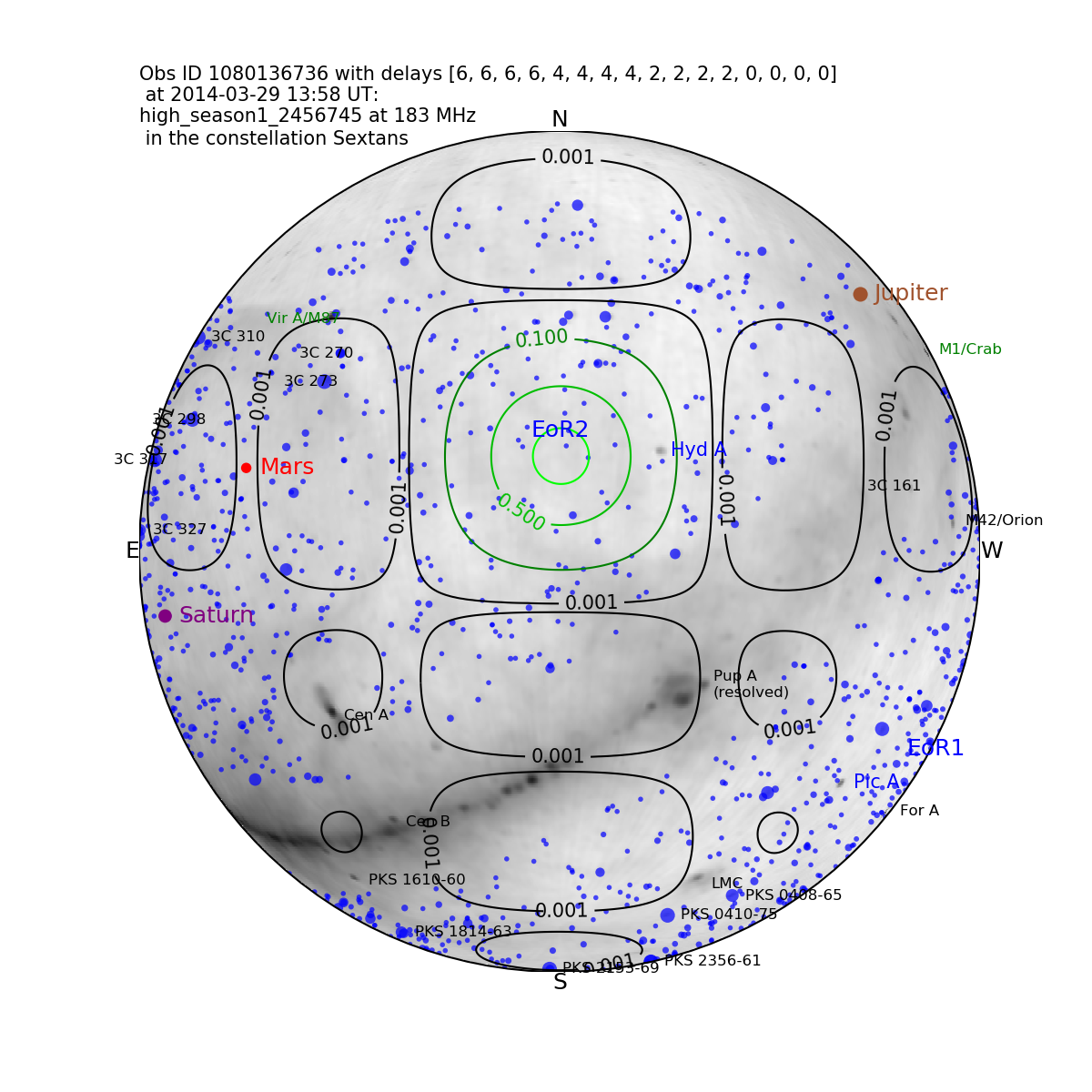}
\includegraphics[width=0.32\textwidth,angle=0]{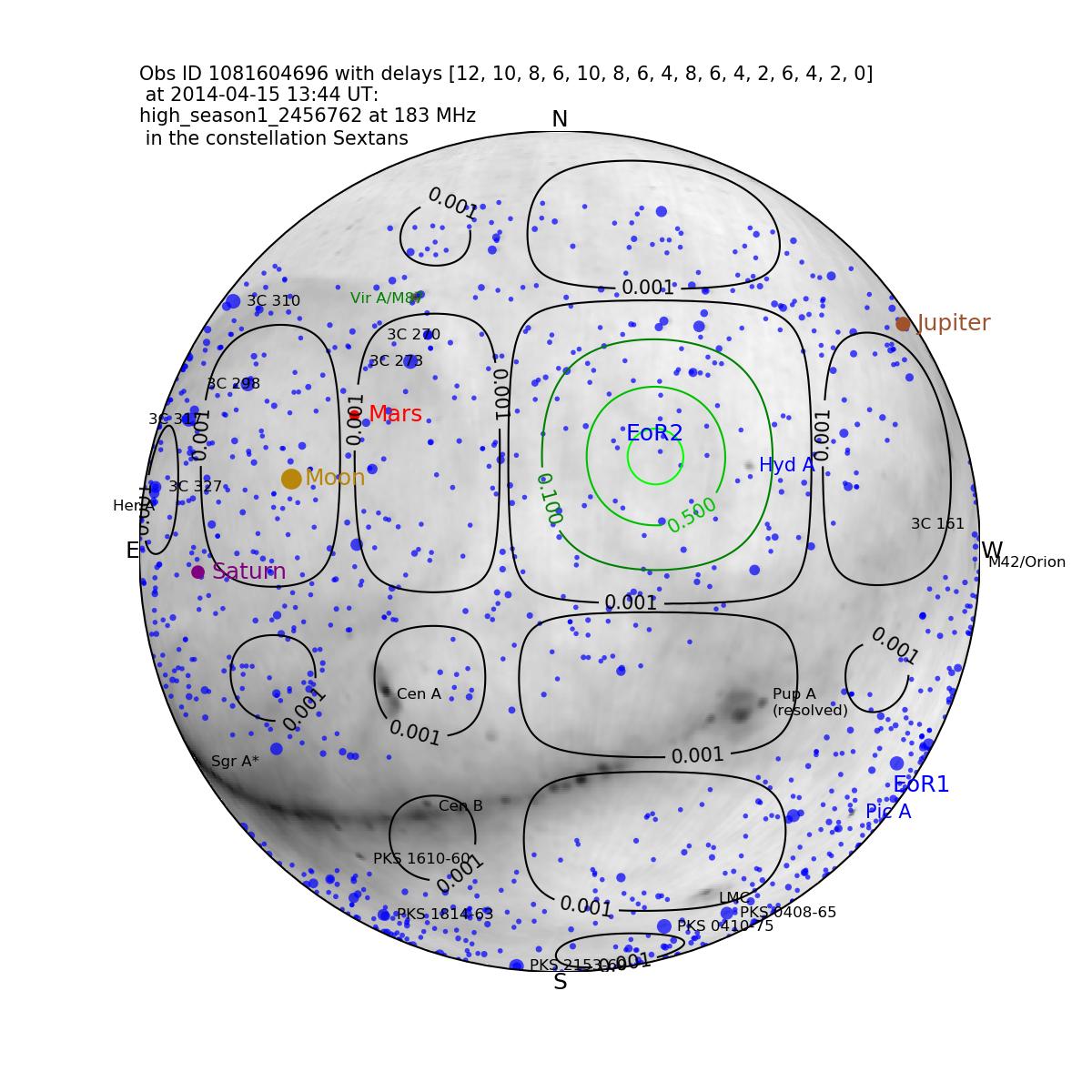}
\caption{Sky response of the telescope for three pointings of the EoR2 field: Minus2 (left), Zenith (centre), Plus2 (right). The rotation of the Galactic plane and Puppis A through the beam sidelobes imprints different structure along the horizon, and on large scales, for the three pointings.}
\label{fig:eor2high_sky}
\end{figure*}

The EoR2 field is centred at RA=10h, Dec.=$-$10 deg. It contains the bright radio galaxy Hydra A on the edge of the main lobe of the primary beam, and the Galactic Plane with the Puppis A supernova remnant and Centaurus A rotating through a 0.1--1.0\% sidelobe. Figure \ref{fig:eor2high_sky} shows the primary beam response at 180~MHz for three pointings used in this work and the EoR2 field.
The Galactic Plane is most prominent in the Plus2 pointing, with structure over a range of spatial scales. Given its location in a sidelobe, we expect its power spectrum signature to imprint power along the horizon line at a range of $k$-modes. The degree of structure in the beam sidelobes will result in time-dependent instrumental spectral indices for these complex sources, and the best outcomes for EoR2 will require subtraction of models for the Galactic Plane and prominent features.

\begin{figure*}
\includegraphics[width=1.\textwidth,angle=180]{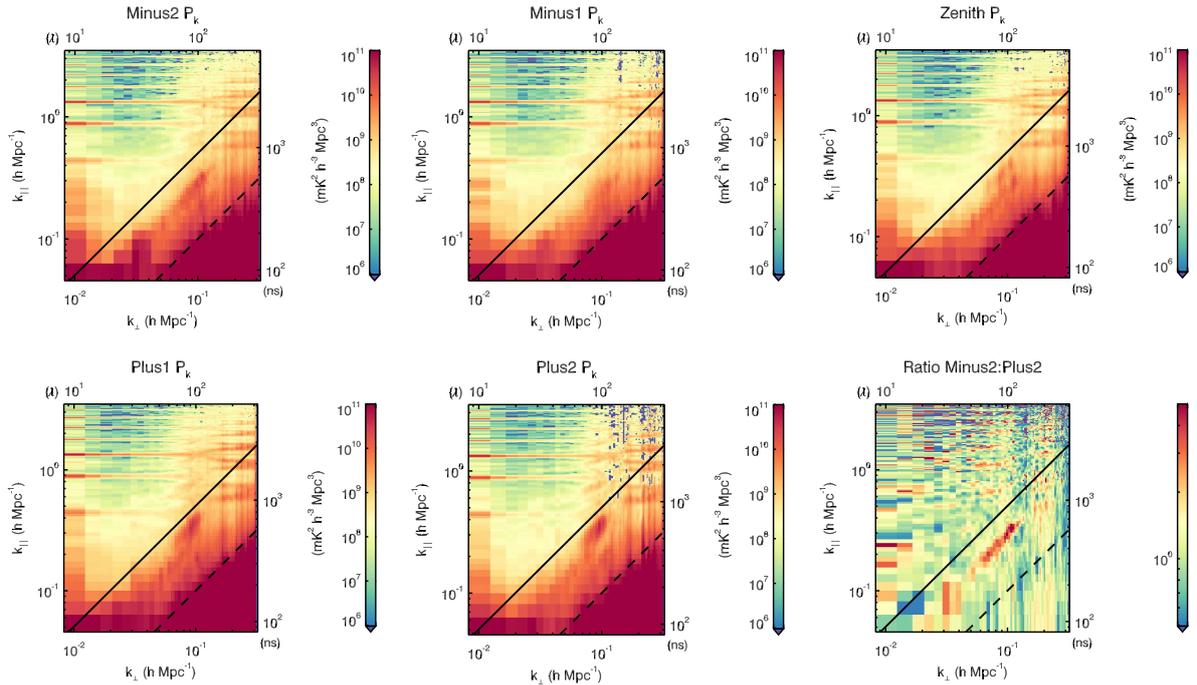}
\caption{Set of power spectra from different pointings for the EoR2 field and full high-band, displaying the variation in primary beam sampling of the residual sky. The rotation of structures in and close to the Galactic Plane through different primary beam responses combined with the evolution of the beam response with frequency, lead to shifting foreground contamination in parameter space.}
\label{fig:eor2high_5pointings}
\end{figure*}

The equivalent 2D power spectra from each pointing, and the ratio of power in the Minus2 to Plus2 pointings, are shown in Figure \ref{fig:eor2high_5pointings}. The aggregated data include 1,420 observations and are therefore representative of the signal in the pointings (i.e., not thermal-noise dominated). The parameter space and colour scale have been reduced to highlight the differences outside of the main foreground-dominated region at low $k_\parallel$.
The data broadly show more contamination in the EoR window ($k_\parallel<0.4$) than was observed with the EoR0 field. This is due to the increased Galactic Plane power in the beam in EoR2, the simple two point-source model used for Hydra A, and the fact that the sky model for EoR0 has received a lot of attention from the collaboration, while EoR2 has been largely ignored (i.e., EoR0 is our primary field). Encouraging results in this work will motivate a better focus of effort on the EoR2 sky model.

\begin{figure}
\includegraphics[width=0.5\textwidth,angle=0]{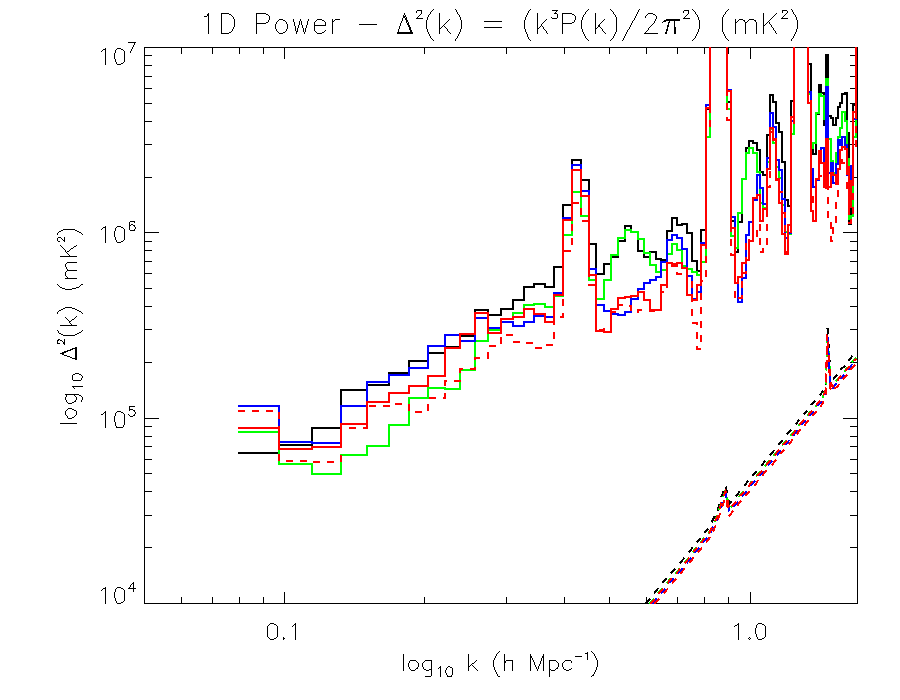}
\caption{Set of 1D power spectra from five different pointings for the EoR2 field and the full high-band; Minus2 (blue), Minus1 (red), Zenith (red-dashed), Plus1 (black), Plus2 (green). Some pointings show improved results at large scale.}
\label{fig:eor2high_pointings_1d}
\end{figure}

Figure \ref{fig:eor2high_pointings_1d} shows the 1D power spectra from each of the five central pointings.
There is some statistically different behaviour from different pointings. This is likely owing to the changing spectral and spatial sampling of the Galactic Plane and major extended radio galaxies. Without careful modelling of this field, which will be explored in coming work, we can only speculate about these differences.

\subsection{Combined results}\label{sec:combined}
Ultimately, the aim is to combine data from individual pointings coherently. The data have been processed using the same phase centre, with a common framework, and with an optimal weighting to allow for coherent addition of data (i.e., the weights are carried through the analysis such that separate datasets can be added using an optimal weighting). However, the primary beam response of the telescope changes appreciably between pointings, and there is scope for decoherence if the telescope response is not correctly modelled, which would lead to undesirable signal loss.

\begin{figure}
\includegraphics[width=0.5\textwidth]{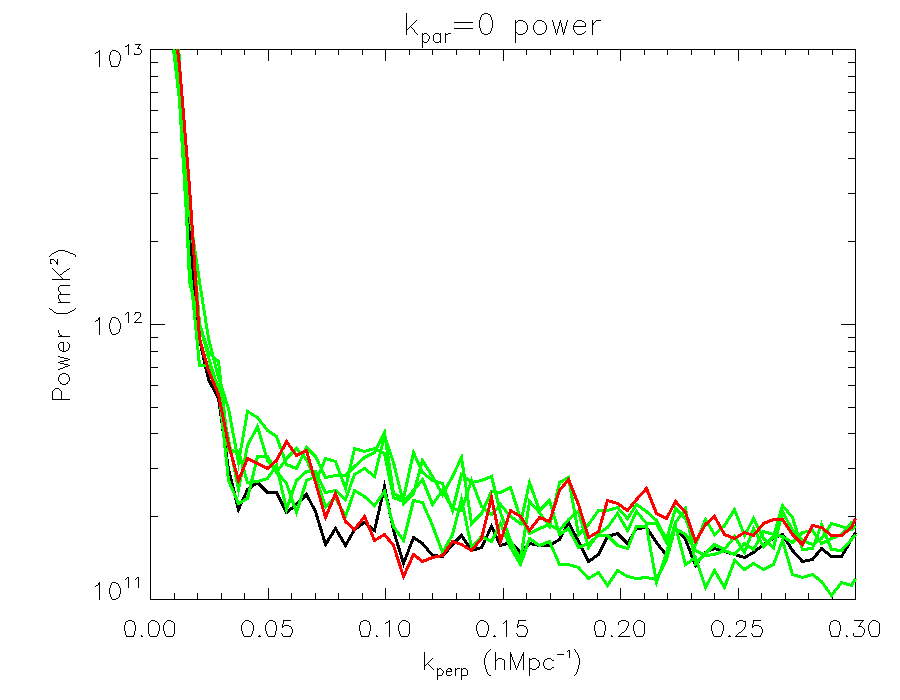}
\caption{Comparison of $k_\parallel =0$ mode power for the individual pointings for EoR0High at $z=6.5$, and the combined power (zenith red; off-zenith green; combined black). The consistency demonstrates that combining pointings coherently is reasonable, because power is not being lost.}
\label{fig:pointingpower}
\end{figure}

\begin{figure}
\includegraphics[width=0.5\textwidth]{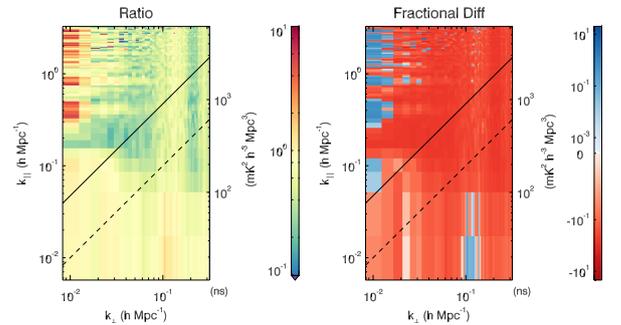}
\caption{Ratio (left) and fractional difference (right) of 2D power for the zenith pointing and combined pointings for EoR0High at $z=6.5$. A ratio consistent with unity and small fractional difference ($<10\%$) in the EoR wedge (low $k_\parallel$) demonstrates that combining pointings coherently is reasonable, because power is not being lost.}
\label{fig:pointingpower2d}
\end{figure}

We can test for this decoherence by comparing the regions of parameter space that should be consistent, e.g., the foreground-dominated EoR Wedge region at low $k_\parallel$. Because this is sky power, it should be retained between pointings and upon coherent addition of pointings. It will not be identical; the different sky response will mean that the power is in different locations, but it should retain an overall power level. Figure \ref{fig:pointingpower} plots the $k_\parallel =0$ mode power for the zenith (red), off-zenith (green) and combined (black) datasets for EoR0High at $z=6.5$. The power is retained during coherent addition. Figure \ref{fig:pointingpower2d} then displays the ratio (left) and fractional difference (right) of 2D power for the zenith pointing and combined pointings. A ratio consistent with unity and small fractional difference ($<10\%$) in the foreground wedge (low $k_\parallel$) demonstrates that combining pointings coherently is reasonable, because power is not being lost.

\begin{figure}
\includegraphics[width=0.5\textwidth]{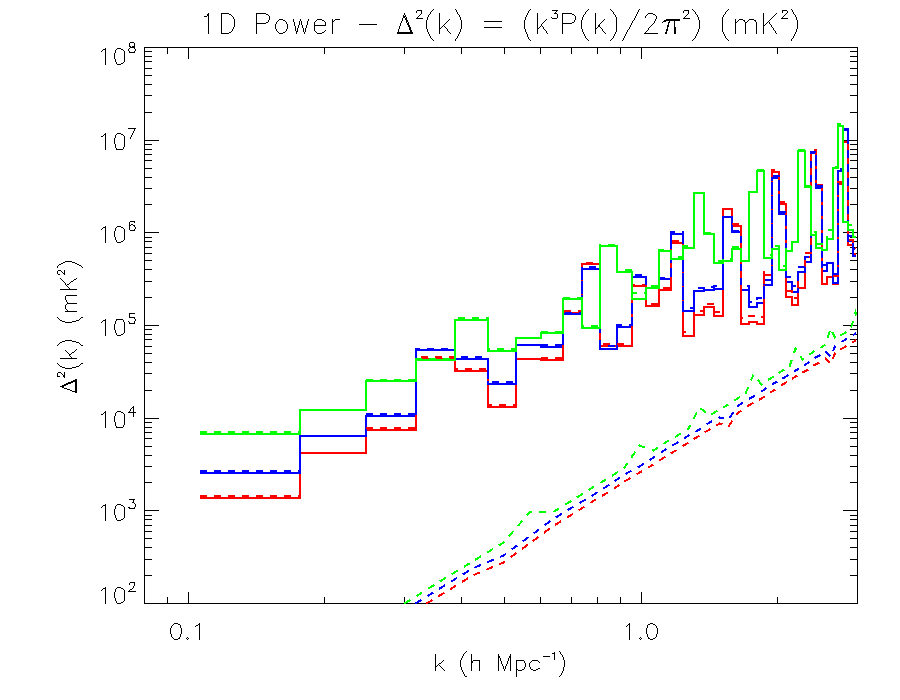}
\caption{Measured 1D power spectrum (solid lines) and measured plus 2$\sigma$ thermal$+$sample variance uncertainty (dashed lines) for the best 3,340 observations (110 hours) from EoR0 high-band at $z=6.5$ (red), mid-band at $z=6.8$ (blue), and low-band at $z=7.1$ (green).}
\label{fig:1d_eor0high_192chan_3340_highmidbot}
\end{figure}

\begin{figure}
\includegraphics[width=0.5\textwidth]{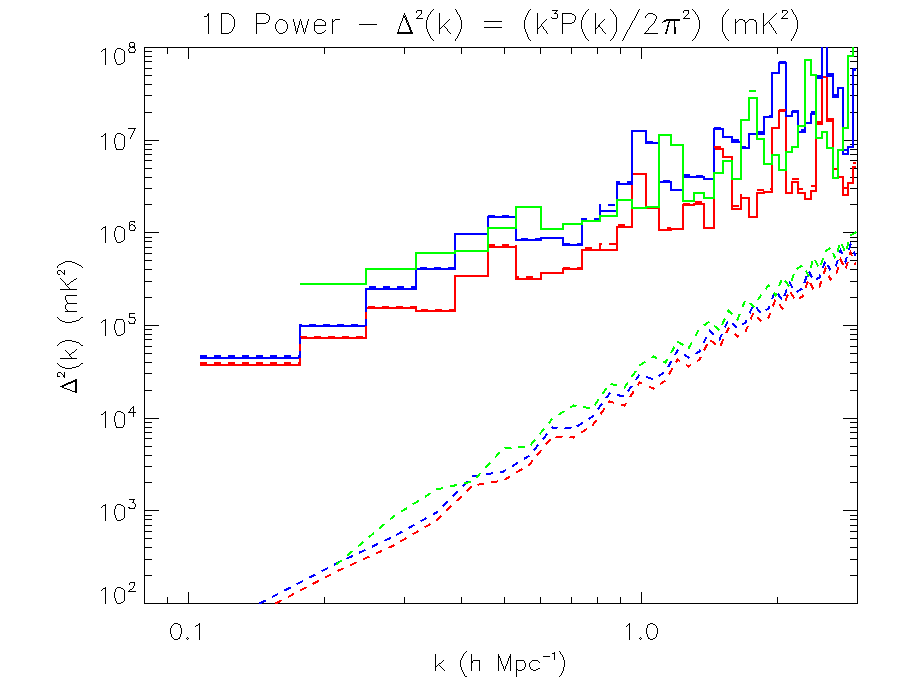}
\caption{Measured 1D power spectrum (solid lines) and measured plus 2$\sigma$ thermal$+$sample variance uncertainty (dashed lines) for the best 1,140 observations (38 hours) from EoR0 low-band at $z=7.8$ (red), mid-band at $z=8.2$ (blue), and low-band at $z=8.7$ (green).}
\label{fig:1d_eor0low_192chan_1140_highmidbot}
\end{figure}

\begin{figure}
\includegraphics[width=0.5\textwidth]{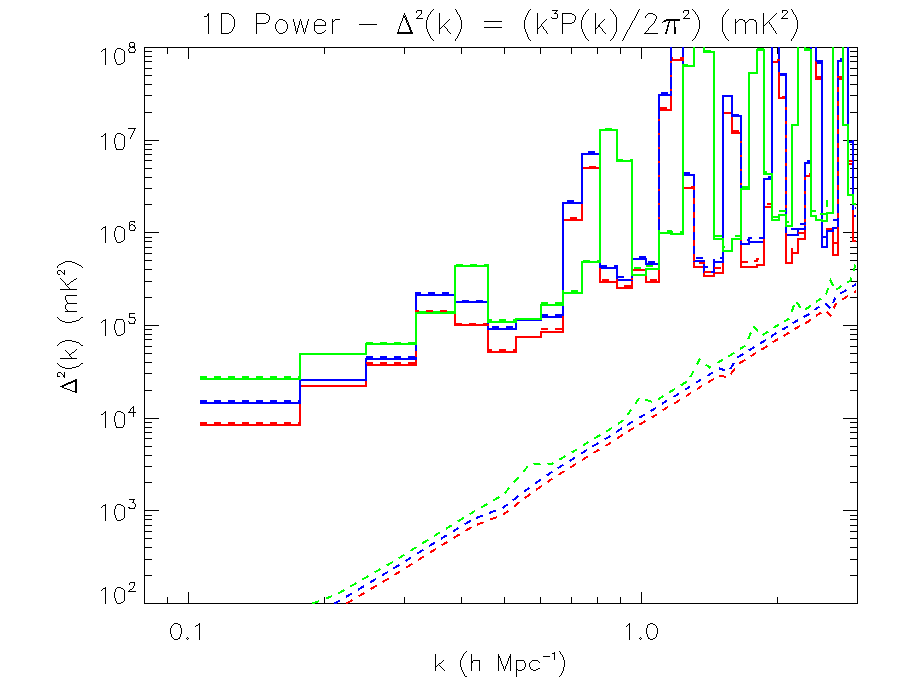}
\caption{Measured 1D power spectrum (solid lines) and measured plus 2$\sigma$ thermal$+$sample variance uncertainty (dashed lines) for the best 1,420 observations (47 hours) from EoR2 high-band at $z=6.5$ (red), mid-band at $z=6.8$ (blue), and low-band at $z=7.1$ (green).}
\label{fig:1d_eor2high_192chan_1420_highmidbot}
\end{figure}

\begin{figure}
\includegraphics[width=0.5\textwidth]{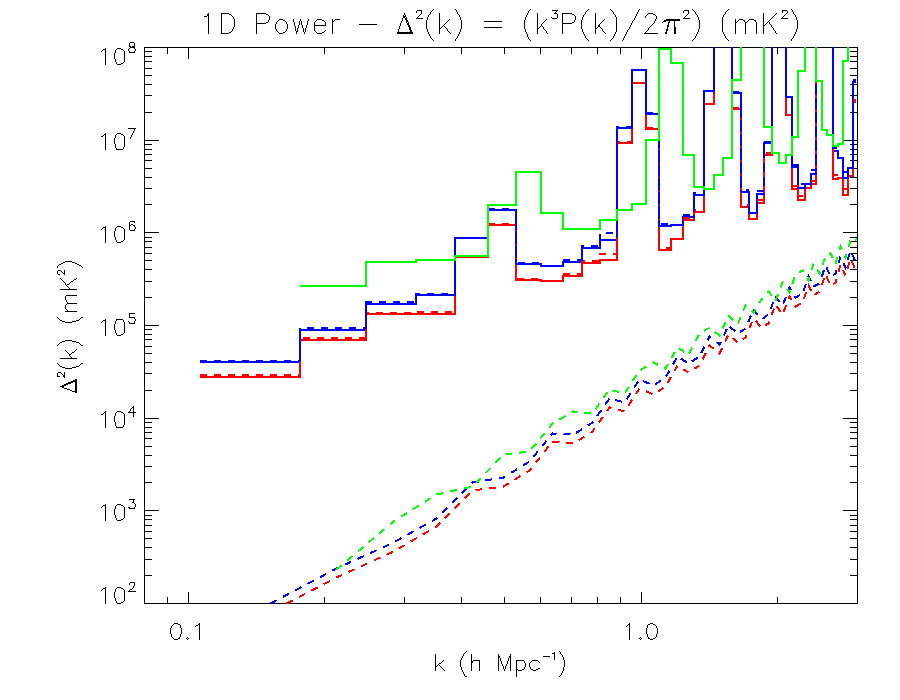}
\caption{Measured 1D power spectrum (solid lines) and measured plus 2$\sigma$ thermal$+$sample variance uncertainty (dashed lines) for the best 1,540 observations (51 hours) from EoR2 low-band at $z=7.8$ (red), mid-band at $z=8.2$ (blue), and low-band at $z=8.7$ (green).}
\label{fig:1d_eor2low_192chan_1540_highmidbot}
\end{figure}

Having established that coherent addition of data from the same observing field and frequency range, is possible, we combine the best zenith-pointed observations with the four off-zenith pointings for EoR0 high- and low-bands, and the EoR2 high- and low-bands. Note that we do not present off-zenith pointings for the EoR1 field, due to the poor results from the zenith data.
Figures \ref{fig:1d_eor0high_192chan_3340_highmidbot} and \ref{fig:1d_eor0low_192chan_1140_highmidbot} display EoR0 results, and Figures \ref{fig:1d_eor2high_192chan_1420_highmidbot} and \ref{fig:1d_eor2low_192chan_1540_highmidbot} display EoR2 results.

\begin{table}{$\Delta$ (mK)}
\centering
\begin{tabular}{|c||c|c|c||c|}
\hline 
Redshift & $k$ ($h$Mpc$^{-1}$) & EoR0 (mK) & EoR1 (mK) & EoR2 (mK) \\
\hline
$z=6.5$ & 0.142 & {\bf 43.1} & 183.8 & 87.1 \\
 & 0.212 & 70.2 & 254.4 & 147.1 \\
 & 0.283 & 93.3 & 403.5 & 189.0 \\
 & 0.354 & 209.5 & 1060.5 & 361.3 \\
 & 0.425 & 183.5 & 876.1 & 305.5\\
 & 0.495 & 125.5 & 455.3 & 232.3 \\
 & 0.566 & 210.1 & 694.7 & 270.7 \\
 & 0.637 & 214.1 & 671.6 & 304.7 \\
 & 0.708 & 384.6 & 1148.5 & 1037.8\\\hline
$z=6.8$ & 0.142 & {\bf 60.1} & 199.9 & 114.3\\
 & 0.212 & 90.0 & 304.2 & 160.6\\
 & 0.283 & 114.1 & 455.7 & 217.7\\
 & 0.354 & 243.9 & 1161.5 & 436.2\\
 & 0.425 & 221.3 & 1024.7 & 407.2\\
 & 0.495 & 169.0 & 541.7 & 323.6\\
 & 0.566 & 255.4 & 840.3 & 327.3\\
 & 0.637 & 260.3 & 842.8 & 340.8\\
 & 0.708 & 383.1 & 1280.9 & 1214.7\\\hline
 $z=7.1$ & 0.142 & {\bf 77.7} & 305.0  & 176.3\\
 & 0.212 & 117.4 & 433.5 & 248.0\\
 & 0.283 & 152.3 & 605.1 & 252.9\\
 & 0.354 & 281.5 & 1111.4 & 434.1\\
 & 0.425 & 263.3 & 1736.6 & 817.0\\
 & 0.495 & 231.9 & 1032.4 & 322.2\\
 & 0.566 & 310.9 & 883.3 & 296.9\\
 & 0.637 & 333.8 & 1001.3 & 410.2\\
 & 0.708 & 437.9 & 1316.2 & 515.2\\\hline
 $z=7.8$ & 0.142 & 229.6 & 571.5 & {\bf 154.2}\\
 & 0.212 & 318.2 & 853.3 & 247.5\\
 & 0.283 & 415.5 & 1119.4 & 314.5\\
 & 0.354 & 417.4 & 1179.6 & 460.1\\
 & 0.425 & 822.2 & 2343.2 & 804.4\\
 & 0.495 & 1146.6 & 3289.9 & 466.8\\
 & 0.566 & 577.4 & 1574.4 & 484.4\\
 & 0.637 & 566.6 & 1436.5 & 501.0\\
 & 0.708 & 667.6 & 1787.7 & 613.4\\\hline
 $z=8.2$ & 0.142 & 223.5 & 787.8 & {\bf 167.7}\\
  & 0.212 & 376.3 & 1166.0 & 430.3\\
 & 0.283 & 421.8 & 1520.2 & 422.2\\
 & 0.354 & 524.2 & 1678.9 & 540.9\\
 & 0.425 & 763.8 & 3102.9 & 772.8\\
 & 0.495 & 1421.0 & 4165.7 & 1402.6\\
 & 0.566 & 981.7 & 2256.5 & 1109.9\\
 & 0.637 & 723.2 & 2112.2 & 739.1\\
 & 0.708 & 719.1 & 2455.4 & 781.1\\\hline
 $z=8.7$ & 0.142 & 353.4 & 1047.3 & {\bf 249.6}\\
  & 0.212 & 544.7 & 1586.2 & 569.9\\
 & 0.283 & 607.9 & 1949.3 & 562.5\\
 & 0.354 & 725.1 & 2087.2 & 688.1\\
 & 0.425 & 826.9 & 3772.3 & 963.2\\
 & 0.495 & 1341.0 & 5214.3 & 1854.5\\
 & 0.566 & 1146.4 & 2754.8 & 1546.0\\
 & 0.637 & 950.7 & 2604.1 & 962.3\\
 & 0.708 & 906.6 & 3078.2 & 947.6\\\hline

\end{tabular}
\caption{Two sigma upper limits on the amplitude of the EoR signal (temperature units: square-root of dimensionless power) for each observing field and redshift. At each redshift, the best limit is bold-faced.}\label{table:results}
\end{table}
The best results at each redshift are reproduced in Table \ref{table:results_best}.
\begin{table}
\centering
\begin{tabular}{|c||c|c|c|}
\hline 
Redshift & $k$ ($h$Mpc$^{-1}$) & UL (mK) & Field \\
\hline
6.5 & 0.142 & 43.1 & EoR0 \\
6.8 & 0.142 & 60.1 & EoR0 \\
7.1 & 0.142 & 77.7 & EoR0 \\
7.8 & 0.142 & 154.2 & EoR2 \\
8.2 & 0.142 & 167.7 & EoR2 \\
8.7 & 0.142 & 249.6 & EoR2 \\
\hline
\end{tabular}
\caption{Best two sigma upper limits on the amplitude of the EoR signal (temperature units: square-root of dimensionless power) for each redshift. }\label{table:results_best}
\end{table}

\subsection{Comparison of fields}\label{sec:comparison}
The results of combining data from different pointings for EoR0 and EoR2 demonstrate better performance in EoR0 at low redshift and EoR2 at high redshift. Given that the distributions of ionospheric activity and EoR Window Power are comparable between the fields, this is likely due to the different Galactic and extended structures drifting through the primary beam sidelobes as a function of frequency. The MWA primary beam introduces strong spectral gradients in the beam nulls, amplifying any effect of mis-modelling of the sources or primary beam in this regions, and potentially imprinting strong spectral structure on residual signals.

\begin{figure}
\includegraphics[width=0.5\textwidth,angle=180]{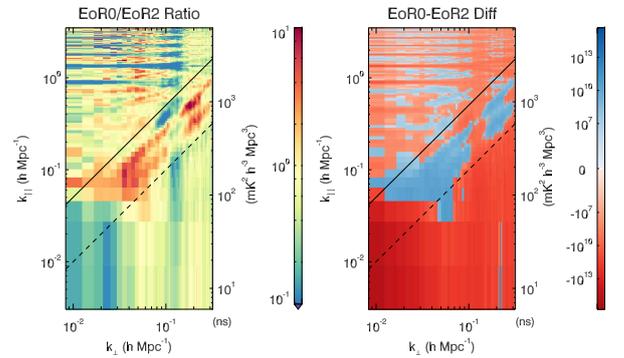}
\caption{Ratio of power in the EoR0 high band to the EoR2 high band, for a similar number of observations. The differing residuals for the two fields are reflected in the differing power structures.}
\label{fig:eor0_eor2_ratiodiff}
\end{figure}

In EoR0, the Galaxy presents more prominently at low frequency due to the larger beam size, whereas the Puppis A, Centaurus A and Centaurus B sources in the EoR2 sidelobes may be better placed with respect to large spectral gradients in the primary beam. Figure \ref{fig:eor0_eor2_ratiodiff} demonstrates this in the high-band 2D power spectrum, showing the ratio and difference of the power in the EoR0 to EoR2 fields. Aside from an overall decrement in the power in EoR0 (red), there is a power increase along the horizon modes in EoR0, indicative of the effect of the spectrally-smooth Galactic Plane. Conversely, the lack of any extended models for the spectrally complex Centaurus A and Puppis A sources leads to additional leaked power in the EoR Window on large scales.

\begin{figure*}
\includegraphics[width=0.49\textwidth,angle=0]{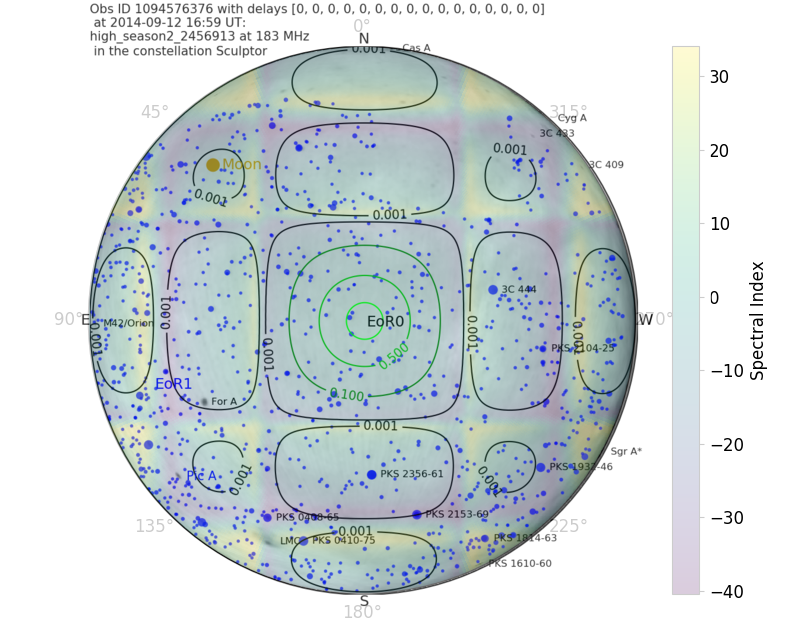}
\includegraphics[width=0.49\textwidth,angle=0]{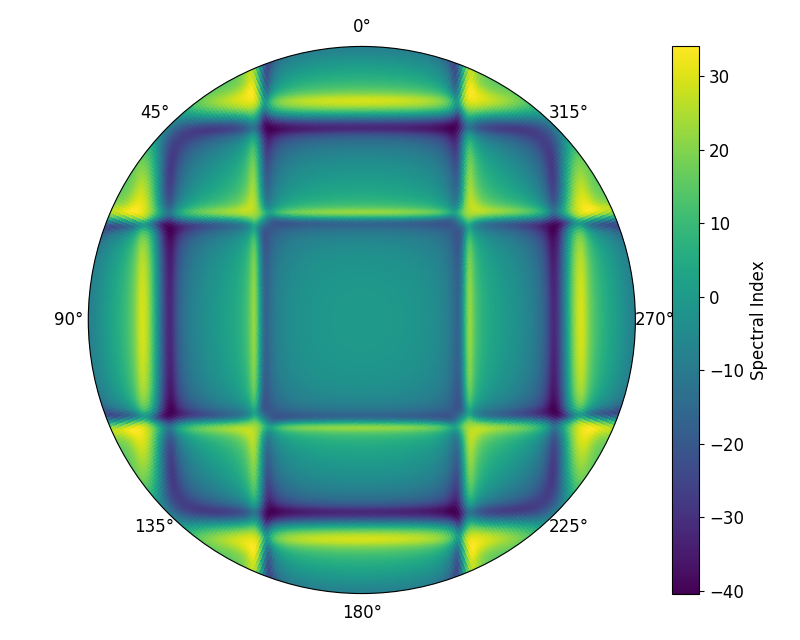}
\caption{Sky response of the telescope for the zenith pointing of the EoR0 high-band (left) and the instrumental spectral index for the same pointing (right). Large instrumental indices imprint spectral structure that can be difficult to remove accurately.}
\label{fig:eor0high_gradient}
\end{figure*}

\begin{figure*}
\includegraphics[width=0.49\textwidth,angle=0]{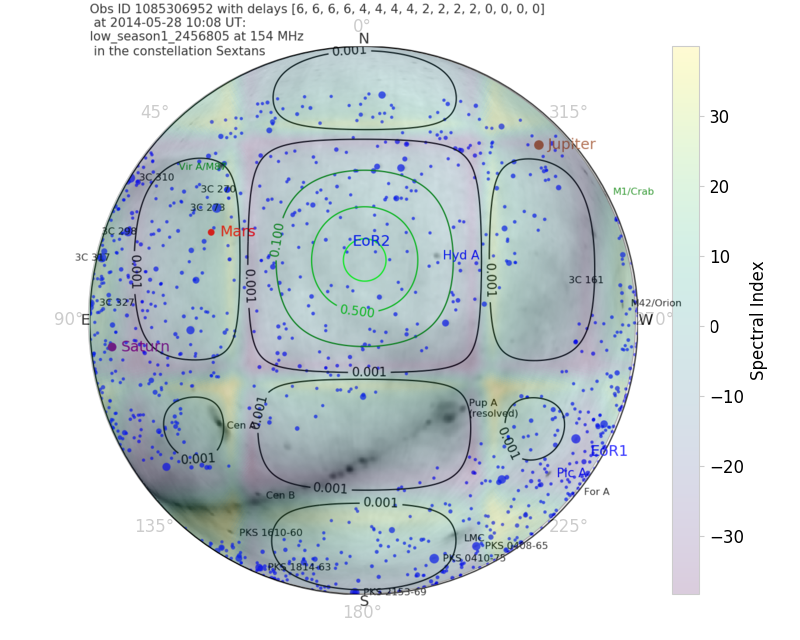}
\includegraphics[width=0.49\textwidth,angle=0]{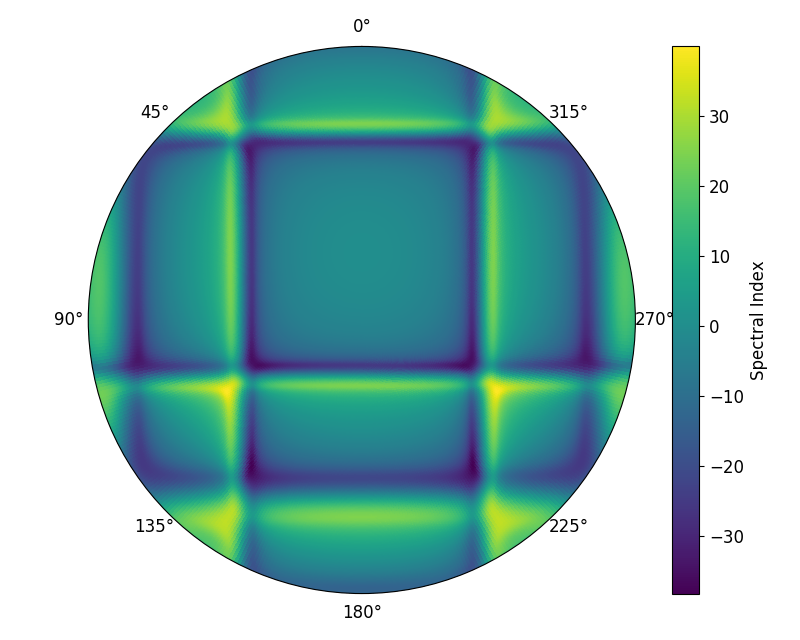}
\caption{Sky response of the telescope for the zenith pointing of the EoR2 low-band (left) and the instrumental spectral index for the same pointing (right). Large instrumental indices imprint spectral structure that can be difficult to remove accurately. The Galactic Plane and extended radio galaxies lie close to spectrally-steep regions of the primary beam.}
\label{fig:eor2low_gradient}
\end{figure*}

We can study the instrumental response to the sky for the pointings where the results are best: EoR0 in high-band, and EoR2 in low-band. In Figures \ref{fig:eor0high_gradient} and \ref{fig:eor2low_gradient} we overlay the sky with the spectral gradient of the primary beam response, for the zenith pointing of EoR0 and EoR2, respectively \citep{cook20}. In the right-hand plots, the beam gradient is shown separately for clarity.

In EoR0, the Galactic Plane is in the spectrally-flat horizon region, and the bright source 3C444 (80~Jy) is away from regions of large index. No bright extended sources reside in regions of large spectral gradient.

In EoR2, the Galactic Plane, with a large number of supernova remnants (e.g., Puppis A, $>$200~Jy), and extended radio galaxies (e.g., Centaurus A, $>$1000~Jy, and Centaurus B, $>$100~Jy) contains most of its structure in spectrally-flat regions, with high spectral indices mostly avoided by these complex sources.

For both of these fields, one can see that small changes in the location of the spectrally-steep beam nulls (as occurs when changing frequency bands) could lead to complex sources incurring large instrumental indices. Given that these bright sources near the field edges are rarely well-subtracted in the current sky models (if a good model even exists), there is potential for leakage into the EoR Window. Dead dipoles (i.e., those where the dipole is present but not delivering signal) within a tile will tend to smooth out these nulls, leading to additional complexity in the sampled signal. The impact of this for these fields is left for future work.

\begin{figure*}
\includegraphics[width=0.49\textwidth,angle=0]{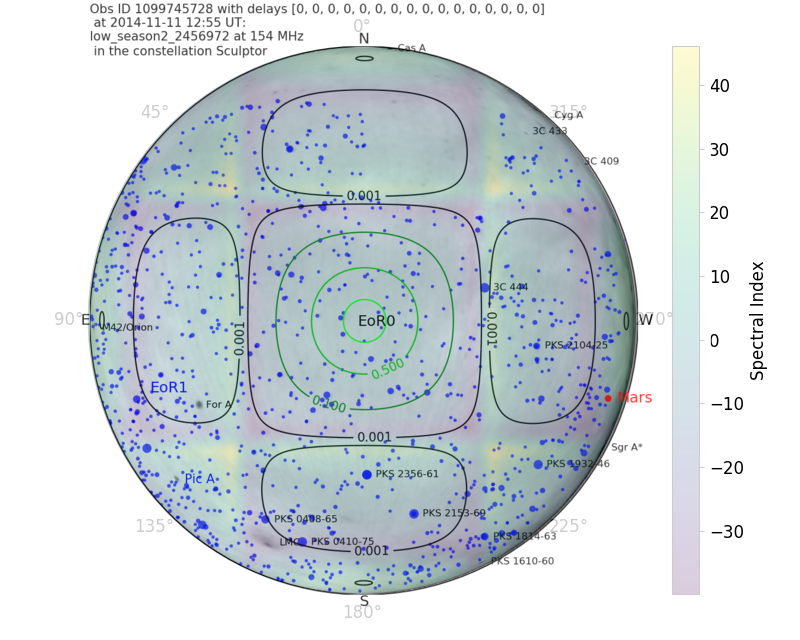}
\includegraphics[width=0.49\textwidth,angle=0]{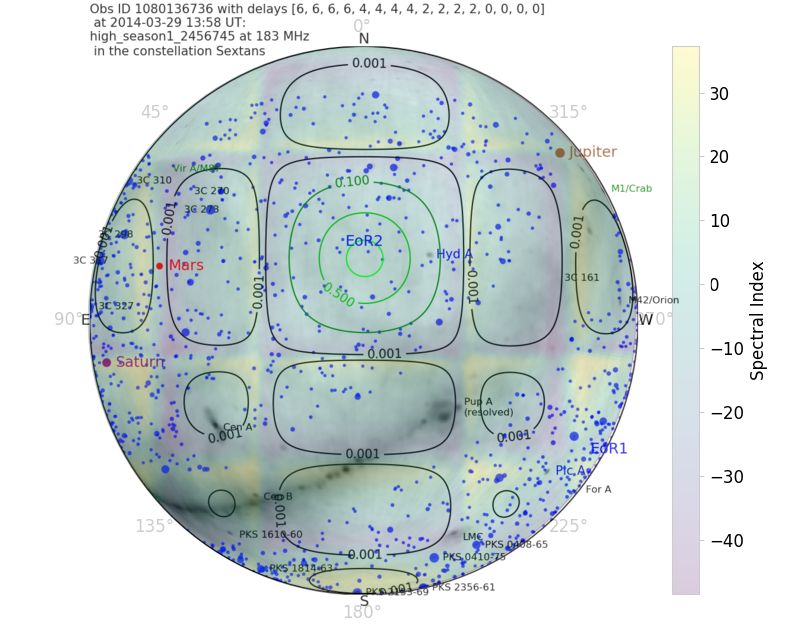}
\caption{Sky response of the telescope for the zenith pointing of the EoR0 low-band (left) and EoR2 high-band (right). 3C444 resides in an area of large spectral index in EoR0. In EoR2, large areas of the Galactic Plane overlay regions of large spectral index.}
\label{fig:eor0eor2_gradient}
\end{figure*}

To study this, we plot the overlaid images for the EoR0 Low-band and EoR2 High-band (zenith) in Figure \ref{fig:eor0eor2_gradient}.
In EoR0, 3C444 (50~Jy) resides in an area of large spectral index. As one of the brightest sources in the field, this may be having an impact if not adequately modelled. In EoR2, large areas of the Galactic Plane overlay regions of large spectral index. Without more in-depth modelling, which we leave to the next stage of this work, it is difficult to ascertain the direct impact of this complicated field.

\section{Discussion and next steps}
This work has presented the most broad and complete census of data from the MWA EoR project over four observing seasons, and multiple fields and redshifts. Despite the same metrics being applied to all datasets, there are clear differences in the structure and power levels of the different fields and observing frequencies. These differences point to sky calibration and source subtraction models as primary drivers, underpinned by the different spatial and spectral structure present in each field. Although EoR1 has only Fornax A as the apparent complex source, the model used for it in this work is sub-optimal. It is difficult to say definitively whether Fornax A residuals are dominating the errors, or whether other components of the sky model are inaccurate.

Conversely, EoR0 and EoR2 show promising results. EoR0 appears clean of bright, extended sources, and has had the most concentrated effort on sky modelling. It produces the best results at low redshift. EoR2 has received the least attention, and contains bright and structured sources in spectrally-steep regions of the instrument response. It shows the best results at high redshift, providing good motivation to improve its sky modelling to further improve results. These lessons can feed into planning for observing fields with the SKA.

The upper limits presented in this work are competitive in the research field at a number of redshifts, particularly when considering the wavemodes ($k$-modes) where the 21cm signal is expected to be strongest. We are confident that improved modelling of the sky in these fields will yield better results from the same underlying datasets. In future work we will use the multi-redshift limits to constrain models of reionisation.

This work comes after the publication of MWA EoR upper limits from 2013 Phase I EoR0 high-band data \citep{beardsley16,barry19a} and 2016 Phase II EoR0 high-band data \citep{li19}. Those works focussed on improved calibration and data analysis strategies to produce excellent results at low redshift. Their $z=6.5-6.8$ results are comparable to those here, although their limits are at a different wavenumber; $k = 0.20, k=0.59~h$Mpc$^{-1}$ compared with $k=0.14~h$Mpc$^{-1}$. It is interesting that their analysis using FHD and $\epsilon$ppsilon yields systematic errors at different wavenumbers than the RTS plus CHIPS pipeline. With 110 hours of data, this work produced upper limits only a factor of 1.5 lower than \citet{li19} produced with 40 hours, showing that systematics are clearly still a dominant factor in our results. The previous work did not have the benefit of the cuts made in this work, but this work also did not fully utilise the calibration and data quality improvements (e.g., SSINS) used in their work. We have confidence therefore that combining these efforts will lead to some further improvement, albeit modest. A larger step to reducing limits can be achieved by updating the MWA signal backend, removing the two-stage filterbank that produces the coarse band structure and forces us to remove spectral channels.

Looking to the future, the lessons learned from studying this broad set of data, combined with previous lessons in the research field, can be applied to future MWA and SKA analyses. We remind the reader that current EoR experiments are systematics-limited, as evident from the results presented here. Given this, from these data and results, the most critical elements for further improvement are:
\begin{itemize}
    \item Bandpass modelling and calibration to accurately remove instrumental chromaticity;
    \item Identification and flagging of low-level RFI in datasets \citep{wilensky19}, as shown in \citet{barry19a} to yield improved results;
    \item Accurate extended source models for bright radio galaxies and Galactic sources, particularly when they reside close to beam nulls \citep{procopio17,line19}.
\end{itemize}
The latter will be achieved through the Long Baseline EoR Survey \citep[LoBES,][]{lynch20}, which has observed the EoR fields, and their flanking fields with Phase II of the MWA, and will form the deepest and most complete low-frequency catalogue in EoR fields.

Conversely, we can also comment on the elements that are not expected to yield large improvements in the results:
\begin{itemize}
    \item Further constraint of ionospheric activity: comparison of datasets with IonoQA =  [2, 5] and IonoQA $>$ 5 yield comparable results;
    \item Different selection of observing fields; these fields are still expected to be the coldest given the size of the MWA primary beam.
\end{itemize}
This latter point has some support from the results of using different amounts of data for the EoR0 high-band (Figure \ref{fig:adddata}), and the lack of correlation found between 1D power spectrum limits and ionospheric activity metric (plot not shown). Although previous work has demonstrated that moderate-strong ionospheric activity can leave residual power in the power spectrum \citep{jordan17,trott17}, the data in this work are selected to be ionospherically quiet, and sorting based on that metric as the primary dimension may not be the optimal approach.

Addressing residual spectral structure from non-smooth bandpass solutions, and mis-modelled bright structures with large instrumental spectral indices, will put us on the path to extend this several hundred hour analysis to a thousand hour analysis.

Looking toward SKA, most of these lessons are still relevant. Ionospheric and RFI conditions will be identical for MWA and SKA due to the common site, and SKA's high sensitivity will demand more stringent quality assurance for these effects. The SKA stations will have randomised antenna locations, smoothing the beams nulls and high instrumental spectral index caused by the MWA's regular aperture array grid. However, each station will have its own custom beam structure, and induced spectral index will play a role for bright sources. Even with the SKA, which has smaller fields-of-view than the MWA, the increased sensitivity will mean that there are few (if any) pointings that do not capture some extended or Galactic emission in the sidelobes. Modelling of these will be crucial for SKA's demanding science goals in the EoR and Cosmic Dawn.


\section*{Acknowledgements}
This research was supported by the Australian Research Council Centre of Excellence for All Sky Astrophysics in 3 Dimensions (ASTRO 3D), through project number CE170100013. CMT is supported by an ARC Future Fellowship under grant FT180100321.
The International Centre for Radio Astronomy Research (ICRAR) is a Joint Venture of Curtin University and The University of Western Australia, funded by the Western Australian State government. KT is partially supported by JSPS KAKENHI Grant Numbers JP15H05896, JP16H05999 and JP17H01110, and Bilateral Joint Research Projects of JSPS.
The MWA Phase II upgrade project was supported by Australian Research Council LIEF grant LE160100031 and the Dunlap Institute for Astronomy and Astrophysics at the University of Toronto.
This scientific work makes use of the Murchison Radio-astronomy Observatory, operated by CSIRO. We acknowledge the Wajarri Yamatji people as the traditional owners of the Observatory site. Support for the operation of the MWA is provided by the Australian Government (NCRIS), under a contract to Curtin University administered by Astronomy Australia Limited. We acknowledge the Pawsey Supercomputing Centre which is supported by the Western Australian and Australian Governments. Data were processed at the Pawsey Supercomputing Centre and at DownUnder GeoSolutions Pty Ltd. Simulations were undertaken on OzStar, Swinburne University.


\bsp	
\label{lastpage}
\end{document}